\documentclass[a4paper]{cas-dc}

\usepackage[authoryear]{natbib}

\usepackage{flushend}
\usepackage{graphicx}
\usepackage{caption}
\usepackage{subcaption}
\usepackage{algorithm}
\usepackage{algpseudocode}
\usepackage[algo2e, ruled,vlined]{algorithm2e}
\usepackage{adjustbox}
\usepackage{multirow}
\usepackage{pifont}
\usepackage{color, colortbl}
\usepackage{hyperref}
\usepackage{float}

\definecolor{Gray}{gray}{0.9}

\newcommand{\cmark}{\ding{51}}
\newcommand{\xmark}{\ding{55}}

\def\tsc#1{\csdef{#1}{\textsc{\lowercase{#1}}\xspace}}
\tsc{WGM}
\tsc{QE}
\tsc{EP}
\tsc{PMS}
\tsc{BEC}
\tsc{DE}

\begin{document}
\let\WriteBookmarks\relax
\def\floatpagepagefraction{1}
\def\textpagefraction{.001}

\shorttitle{End-to-End Integration of Speech Separation and Voice Activity\\Detection for Low-Latency Diarization of Telephone Conversations}

\shortauthors{Morrone et~al.}

\title [mode = title]{End-to-End Integration of Speech Separation and Voice Activity Detection for Low-Latency Diarization of Telephone Conversations}




%
\author[1, 2]{Giovanni Morrone}[type=editor,orcid=0000-0003-2163-1779]

\cormark[1]


\ead{g.morrone@almawave.it}



\address[1]{Università Politecnica delle Marche, Ancona, Italy}

\author[1]{Samuele Cornell}

\ead{s.cornell@pm.univpm.it}

\author[1]{Luca Serafini}

\ead{l.serafini@univpm.it}

\author[2]{Enrico Zovato}

\ead{e.zovato@almawave.it}

\address[2]{Almawave S.p.A., Rome, Italy}

\author[3]{Alessio Brutti}

\ead{brutti@fbk.eu}

\address[3]{Fondazione Bruno Kessler, Trento, Italy}

\author[1]{Stefano Squartini}

\ead{s.squartini@univpm.it}

\cortext[cor1]{Corresponding author}



\begin{abstract}
Recent works show that speech separation guided diarization (SSGD) is an increasingly promising direction, mainly thanks to the recent progress in speech separation. 
It performs diarization by first separating the speakers and then applying voice activity detection (VAD) on each separated stream. 
In this work we conduct an in-depth study of SSGD in the conversational telephone speech (CTS) domain, focusing mainly on low-latency streaming diarization applications.
We consider three state-of-the-art speech separation (SSep) algorithms and study their performance both in online and offline scenarios, considering non-causal and causal implementations as well as continuous SSep (CSS) windowed inference.
We compare different SSGD algorithms on two widely used CTS datasets: CALLHOME and Fisher Corpus (Part 1 and 2) and evaluate both separation and diarization performance. 
To improve performance, a novel, causal and computationally efficient leakage removal algorithm is proposed, which significantly decreases false alarms. 
We also explore, for the first time, fully end-to-end SSGD integration between SSep and VAD modules. Crucially, this enables fine-tuning on real-world data for which oracle speakers sources are not available. 
In particular, our best model achieves $8.8$\% DER on CALLHOME, which outperforms the current state-of-the-art end-to-end neural diarization model, despite being trained on an order of magnitude less data and having significantly lower latency, i.e., $0.1$ vs. $1$ seconds.
Finally, we also show that the separated signals can be readily used also for automatic speech recognition, reaching performance close to using oracle sources in some configurations. 

\end{abstract}



\begin{keywords}
online speaker diarization \sep speech separation \sep end-to-end learning \sep overlapped speech \sep conversational telephone speech
\end{keywords}

\maketitle

\section{Introduction}
\label{sec:intro}
Speaker diarization consists in identifying ``who spoke when'' in an input audio, by segmenting it into speaker-attributed regions~\citep{anguera2012speaker,park2022review, serafini2023experimental} that correspond to speakers' utterances. 
It is an essential pre-processing task in many applications, such as lectures, meetings, live captioning, speaker-based indexing, telephone calls, and doctor-patient conversations.

Historically, diarization has relied on clustering-based methods, which have been widely investigated since the 90s, and have represented the de-facto standard approach to diarization for many years. 
Recently, the invention of end-to-end neural diarization (EEND)~\citep{fujita2020end} has shown promising improvements in diarization accuracy, especially in the presence of overlapped speech, which may constitute up to 20\% of total speech in real conversations \citep{watanabe2020chime}. Indeed, classical, clustering-based systems are not able to handle overlapped speech as clustering is usually applied on single-speaker embeddings extracted from short frames. In this case, overlap-aware diarization can be performed using post-processing strategies~\citep{bullock2020overlap, raj2021multi}.
There also exist methods such as target-speaker voice activity detection (VAD)~\citep{Medennikov2020TargetSpeakerVA}, region proposal networks~\citep{huang2020speaker} and speech separation guided diarization (SSGD)~\citep{fang2021deep} that do not belong to these two categories. In particular, SSGD performs diarization by combining speech separation (SSep) and VAD. It is particularly appealing as separated sources could be readily fed in input to an automatic speech recognition (ASR) system (see Section~\ref{subsec:res/wer_eval}).

The majority of the methods above only work offline and thus are not suitable for streaming processing. The extension from offline to online processing is way simpler for EEND. Clustering-based systems consist of a pipeline of several modules (i.e., VAD, speaker embedding extraction, clustering, etc). Each of these modules has to be updated in order to work online. In contrast, end-to-end methods can be easily adapted for online diarization by employing a speaker-tracing buffer (STB) to store previous input-output pairs that can be exploited for online inference~\citep{xue2021edaonline, xue2021online, horiguchi2022online}. An alternative strategy is to simply replace the neural architecture with another one that allows low-latency streaming processing~\citep{han2021bw}. A similar approach, which is the basis for this work, is proposed by \cite{morrone2022low-latency} also for SSGD.

In this preliminary work, SSep and VAD models were trained independently and combined together with no additional training. Despite its simplicity, this approach only reaches a sub-optimal diarization performance. Additionally, a major disadvantage is that the SSep module needs a dataset in which oracle sources for the speakers are available. This could lead to mismatched training-inference conditions as in most real-world scenarios oracle speaker sources are difficult to obtain~\citep{subakan2022real}.
In this paper, we build upon our previous work \citep{morrone2022low-latency} and propose a fully end-to-end integration of the SSep and VAD modules. This approach only requires diarization labels during the fine-tuning stage and thus eliminates the need for oracle speaker target sources.
We focus on the conversational telephone speech (CTS) domain, where the maximum number of speakers is limited to $2$. Despite this limitation, this scenario is quite common in many commercial applications (e.g., doctor-patient recordings).
This choice also allows to compare directly with previous works on EEND diarization.
We show that the proposed end-to-end strategy provides significant improvements on two widely used datasets, i.e., Fisher Corpus \citep{cieri2004fisher} and CALLHOME \citep{callhome}. These reflect two different training vs. inference conditions: fully matched and unmatched respectively. 
In particular, our best online model outperforms the current state-of-the-art EEND system on the $2$-speaker subset of the CALLHOME dataset despite having an order of magnitude lower latency, i.e., $0.1$ vs $1$ seconds.

The main contributions with respect to our previous work \citep{morrone2022low-latency} are summarized below:

\begin{itemize}
\item We propose a joint fine-tuning strategy of SSep and VAD modules which consistently improves over disjoint trained SSGD. Notably, the proposed approach allows fine-tuning on datasets for which separated oracle sources are not available.
\item We thoroughly study the effect of the proposed leakage removal algorithm on both disjoint and end-to-end trained SSGD. Additionally, we analyze the relationship between the leakage removal aggressiveness and the diarization evaluation metrics.
\item We carry out an analysis of the effect of varying model latency on diarization performance.
\item We consider an additional more performing SSep model, i.e., DPTNet \citep{dptnet}. 
\item Since separated sources are not provided in the CALLHOME dataset, we create a simulated version that enables the adaptation of SSep models that further reduces diarization errors.
\end{itemize}

Although it is straightforward to cascade causal SSep and VAD modules to achieve low latency, our experiments show that simple concatenation does not lead to optimal performance. Firstly, the choice of appropriate SSep and VAD models is crucial. Moreover, the use of the leakage removal algorithm consistently improves accuracy for all proposed methods at a very negligible cost. Finally, joint fine-tuning further reduces diarization errors and it is particularly appealing when oracle separated sources are not available (e.g., CALLHOME). We complete our work providing interesting in-depth analysis on several SSGD aspects (e.g., online/offline systems comparison, latency analysis, leakage removal impact, ASR evaluation). Such analysis aims at shedding more light on the pros and cons of using speech separation for the diarization task.

In the next subsection we report a brief summary of the diarization state-of-the-art. In Section \ref{sec:sys} we provide a description of the SSGD framework. The experimental setup is shown in Section \ref{sec:exp_setup}. Experiments and results are reported in Section \ref{sec:res}. Finally, we draw the conclusions in Section \ref{sec:conc}.

\subsection{Related Works}
\label{ssec:intro/rel}
The conventional clustering-based diarization approach is typically a cascade of three tasks: voice activity detection, speaker embedding extraction from speech segments and clustering of the embeddings. Previous works constantly improve the overall performance by developing better methods for one or more tasks. Several papers focus on the development of better speaker embeddings extractors \citep{dehak2010front, garcia2017speaker, desplanques2020ecapa, Xiao2021MicrosoftSD, koluguri2022titanet}. In contrast, in \cite{park2019auto}, \cite{singh2021self} and \cite{landini2022bayesian} different clustering algorithms are proposed. Conventional clustering algorithms can only handle correctly single-speaker segments, thus overlapped speech is usually missed out or incorrectly labeled. To deal with overlapped speech, recent works extend the standard pipeline with overlap assignment techniques \citep{bullock2020overlap, raj2021multi, jung21_interspeech}. These methods need accurate overlap detection, which is often hard to train. Furthermore, embedding extractors trained on single-speaker utterances may not be reliable for overlapping segments, resulting in speaker confusion errors~\citep{raj2021multi}. 

Recently, deep learning methods are employed for end-to-end neural diarization approaches. A major advantage of EEND is that they are able to deal with overlapped speech without any modification. The first EEND methods perform diarization as a simple multi-speaker voice activity detection problem, in which each output represents a different speaker's speech activity. EEND systems can employ different neural architectures, such as bidirectional long-short memory \citep{fujita2019} and self-attention (SA-EEND) \citep{fujita2019end}. EEND-based systems are trained directly to perform diarization using permutation invariant training (PIT) \citep{kolbaek2017multitalker} as the diarization problem is inherently permutation-invariant without any a-priori information. 
With enough training data, end-to-end approaches have been shown to outperform current state-of-the-art clustering-based systems~\citep{Horiguchi2021TheHD}. Contrary to clustering-based approaches, the maximum number of total speakers is fixed in the aforementioned EEND architectures. Additionally, end-to-end systems need to process the entire input signal during inference, resulting in significant memory consumption for long recordings (e.g., >10 minutes). Chunk-wise processing can help but is not viable as it leads to speaker label permutation across chunks due to PIT. \cite{horiguchi2020end} solves the first problem by extending the basic EEND with an auto-regressive encoder-decoder (EEND-EDA) architecture.
The second issue is generally addressed by combining clustering and EEND. In \cite{horiguchi2021end} EEND is exploited to refine the results of a clustering-based algorithm. \cite{bredin2021end} proposes a speaker segmentation model inspired by EEND to perform overlap-aware resegmentation in a conventional diarization pipeline. However, these approaches result in more complex systems in contrast to the simplicity of fully end-to-end architectures. A tighter integration of EEND and clustering, i.e., EEND-vector clustering (EEND-VC), is proposed in \cite{kinoshita21_interspeech, kinoshita2021integrating}. It performs chunk-wise processing to ensure that a given maximum number of active speakers can be present in each chunk (e.g., 2 or 3). The original EEND is modified to output global speaker embeddings that are aggregated across chunks using a constrained clustering algorithm. This method can both deal with an arbitrary number of speakers and solve the inter-chunk speaker permutation problem. An approach similar to EEND-VC, named EEND with global and local attractors (EEND-GLA), is proposed in \cite{horiguchi2021towards} which combines EEND-EDA and unsupervised clustering to deal with cases where the number of speakers appearing during inference is higher than that during training.
Another system \citep{zeghidour2021dive} employs a different architecture that iteratively builds embeddings for each speaker which are exploited to condition a VAD module.

An alternative framework to deal with overlapped speech is continuous speech separation (CSS)~\citep{chen2020continuous, morrone2022conversational}.
CSS extends PIT-based SSep to long recording scenarios, by applying separation in a chunk-wise manner, where each chunk is assumed to contain a fixed number of speakers (usually 2-3). Since the underlying separator is trained via a PIT objective, output permutation consistency between chunks is not guaranteed. CSS solves this problem by performing overlapping inference (i.e., using strides shorter than chunk sizes) and reordering adjacent chunks based on a similarity measure over the portion in which they overlap.  
Several recent works have proposed diarization systems inspired by CSS. In ~\cite{raj2021integration} and \cite{Xiao2021MicrosoftSD} separation is done in windowed segments. In this case, speaker permutation is addressed by applying a diarization method (e.g., clustering-based) across the separated audio streams.
On the other hand, \cite{fang2021deep} proposes SSGD, where diarization is performed by first separating the input mixture and applying a conventional VAD to detect speech segments in each channel. In \cite{morrone2022low-latency}, we improve the SSGD architecture using a neural-based VAD and a novel post-processing algorithm that removes the channel leakage generated by separation. In addition, SSGD is adapted to allow online inference employing causal SSep and VAD models.

All the aforementioned approaches are not suitable for streaming applications (e.g., live captioning) as they only work offline.
Although one could potentially leverage CSS-based systems to allow online inference (as done in \cite{yoshioka2019low}), no established solutions that exploit low-latency SSep for diarization are available in literature, with the exception of \cite{morrone2022low-latency}. Specifically, online inference is not trivial for systems that integrate CSS and clustering-based diarization \citep{raj2021integration, Xiao2021MicrosoftSD} as all submodules have to work online.
Several methods have been proposed to allow EEND systems to deal with online processing. \cite{xue2021online} extends the SA-EEND with an STB mechanism. Inference is performed on short chunks and the speaker permutation information is selected from input-output pairs of previous frames and stored in a buffer. The permutation ambiguity is solved using the buffered frames to condition the output of the current chunk. \cite{xue2021edaonline} and \cite{horiguchi2022online} propose similar STB-based extensions for EEND-EDA and EEND-GLA, respectively. Instead, \cite{han2021bw} improves the online EEND-EDA using chunk-level recurrence to process the chunk hidden states making the model complexity linear in time.
\cite{coria2021overlap} designs a different approach that combines the use of EEND with an x-vector extractor and online clustering. The EEND model is used to gate the representation before the x-vector statistical pooling layer, to extract per-speaker embeddings even in overlap regions.

\section{SSGD Framework}
\label{sec:sys}

\begin{figure}[t]
\centering
\includegraphics[width=0.48\textwidth]{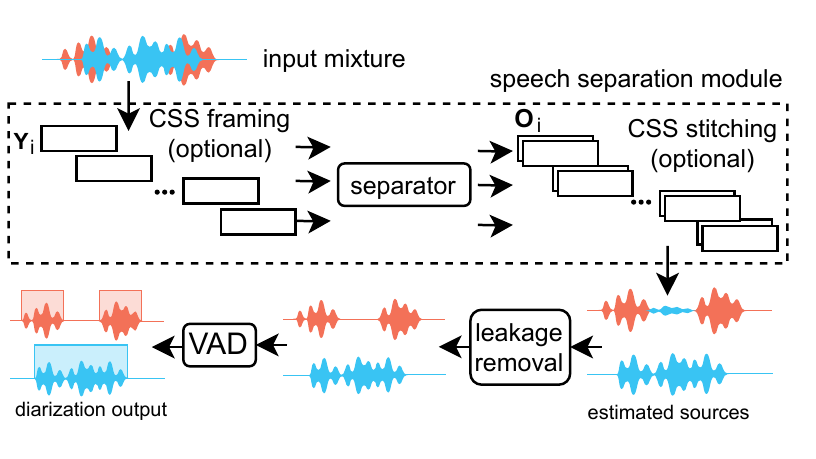}
\caption{General diagram of the SSGD method.}
\label{fig:cssgd}
\end{figure}

Our SSGD pipeline is shown in Fig. \ref{fig:cssgd} and consists of three modules: speech separation, voice activity detection and leakage removal. The system is fed with a single-channel mixed audio input, denoted $\mathbf{Y} \in \mathbb{R}^{1 \times T}$, where $T$ is the number of audio samples.

\subsection{Speech Separation Module}
\label{subsec:sys/css}

In our experiments we employ causal separation models as SSep modules (i.e., Conv-TasNet \citep{luo2019conv}, DPTNet \citep{dptnet} and DPRNN \citep{luo2020dual}).
To compare the proposed SSGD with both clustering-based and EEND state-of-the-art offline systems, we also experiment with non-causal SSep models.

Low latency is achieved in different ways according to the SSep architectures.
For Conv-TasNet, we only use causal convolutions and global layer normalization is replaced with cumulative layer normalization.
For DPRNN, online processing is achieved using LSTM in place of bi-directional LSTM for inter-chunk processing to tie the latency to the intra-chunk segment size \citep{li2023design}.
Regarding DPTNet, a streaming modified version is attained in this work by simply masking the future frames in each self-attention layer in the inter-processing blocks, with the rest being equal to causal DPRNN (DPTNet is identical to DPRNN except for self-attention layers).

Additionally, we consider an alternative approach that uses non-causal SSep models in the CSS framework to allow streaming inference (as done in \cite{chen2020continuous} and \cite{morrone2022conversational}).
In such a configuration, the latency of non-causal models is tied to the CSS window size.
We do not apply CSS in causal SSep models as they do not require chunk-wise processing to work online.

CSS is composed of three stages: framing, separation and stitching. In the framing stage, a windowing operation splits $\mathbf{Y}$ into $I$ overlapped frames $\mathbf{Y}_i \in \mathbb{R}^{1 \times W}, i = 1, \dots, I$, with $I = \lceil \frac{T}{H} \rceil$, where $W$ and $H$ are the window and hop sizes, respectively.
Each frame $\mathbf{Y}_i$ is fed to the separator, which generates $C$ separated output frames $\mathbf{O}_i \in \mathbb{R}^{C \times W}$. $C$ is the maximum number of speakers in each frame. In this work, $C$ is fixed to $2$ as it is a common assumption for telephone conversations.
The output channels could be misaligned due to permutation-free training. The stitching module solves the permutation ambiguity by aligning the channels of two consecutive separated outputs $\mathbf{O}_i$ and $\mathbf{O}_{i+1}$. The correct alignment is estimated according to the cross-correlation between the overlapped portion of the consecutive frames. Finally, the aligned outputs are merged into the final output stream $\mathbf{X} \in \mathbb{R}^{C \times T}$ by using an overlap-add method with Hanning window.

\subsection{Voice Activity Detection Module}
\label{subsec:sys/vad}

The VAD module is fed with the estimated separated sources and detects active speech segments. VAD is applied on each separated source $\mathbf{\tilde{X}_{\ell}}$ independently and the outputs are combined to produce the diarization output.
We investigate the use of two different models: an energy-based conventional VAD ~\citep{landini2021analysis} and a neural-based VAD which employs a temporal convolutional network (TCN) \citep{bai2018empirical}, as proposed in \cite{cornell2022overlapped}. The first method does not require additional training, whilst the latter is data driven.

\subsection{Leakage Removal Module}
\label{subsec:sys/post_proc}

State-of-the-art SSep models have reached impressive performance when tested on short fully overlapped utterances \citep{wang2018supervised}. However, such models could generate channel leakage in sparsely overlapped conversational speech when only one speaker is active \citep{Xiao2021MicrosoftSD}. Thus, the VAD module can detect as speech the ''leaked'' segments. This negatively affects the diarization output by introducing false alarm errors. We propose a lightweight post-processing algorithm to mitigate this problem. It does not introduce additional latency and has very little impact on separation quality, missed speech and speaker confusion errors.  

The leakage removal post-processing algorithm is summarized in Algorithm \ref{algo:leak_rem}.
The algorithm is fed with an input mixture $\mathbf{Y}$ and two estimated sources $\mathbf{X}^1$, $\mathbf{X}^2$ which are split into disjoint segments $\mathbf{Y}_{\ell}$, $\mathbf{X}_{\ell}^1$, $\mathbf{X}_{\ell}^2$ of length $L$.
For each segment, we compute the Scale-Invariant Signal-to-Distortion Ratios (SI-SDR) \citep{leroux2019sdr} $s_{\ell}^1$, $s_{\ell}^2$ between segments of every source $\mathbf{X}_{\ell}^1$, $\mathbf{X}_{\ell}^2$ with the associated segment $\mathbf{Y}_{\ell}$ of input mixture.
A leaked segment is detected when both $s_{\ell}^1$, $s_{\ell}^2$ are above a threshold $t_{\ell r}$. In the SSGD framework leakage is removed by filling with zeros the segments with lower SI-SDR.
Then, the post-processed estimated sources $\mathbf{\tilde{X}_{\ell}}$ are passed as input to the following VAD module.

\SetKwInput{KwInput}{Input}                
\SetKwInput{KwOutput}{Output}
\SetKw{KwBy}{by}
\SetKw{KwAnd}{and}
\begin{algorithm}[H]
\DontPrintSemicolon
  \KwInput{$\mathbf{Y}$, $\mathbf{X}^1$, $\mathbf{X}^2$, $T$, $L$, $t_{\ell r}$}
  \KwOutput{$\mathbf{\tilde{X}_{\ell}}^1$, $\mathbf{\tilde{X}_{\ell}}^2$}
  
  $\mathbf{\tilde{X}_{\ell}}^1 \gets \mathbf{X}^1$; $\mathbf{\tilde{X}_{\ell}}^2 \gets \mathbf{X}^2$ 
  
  
  \For{$i \gets 0$ \KwTo $T$ \KwBy $L$}{
    $s_{\ell}^1 \gets $ SI-SDR$(\mathbf{Y}$[$i$:$i$+$L$], $\mathbf{X}^1$[$i$:$i$+$L$])
    
    $s_{\ell}^2 \gets $ SI-SDR$(\mathbf{Y}$[$i$:$i$+$L$], $\mathbf{X}^2$[$i$:$i$+$L$])
    
    \If{ $s_{\ell}^1 > t_{\ell r}$ \KwAnd $s_{\ell}^2 > t_{\ell r}$}{
      \If{$s_{\ell}^1 > s_{\ell}^2$}{
        $\mathbf{\tilde{X}_{\ell}}^2$[$i$:$i$+$L$] $ \gets 0$
      }
      \Else{
       $\mathbf{\tilde{X}_{\ell}}^1$[$i$:$i$+$L$] $ \gets 0$
      }
    }
    }
\caption{Leakage Removal}
\label{algo:leak_rem}
\end{algorithm}

\subsection{End-to-End Training}
\label{subsec:sys/eend}
In our previous work \citep{morrone2022low-latency} diarization is performed by cascading the SSep, leakage removal and VAD modules. In such a case, the SSep and VAD models are trained independently. We refer to this configuration as \emph{disjoint SSGD}.

In the \emph{end-to-end SSGD} SSep and VAD are jointly optimized, while disabling the leakage removal during training. The SSep and VAD modules are initialized with the parameters of the disjoint SSGD and fine-tuned following two methods. In the first approach, i.e., \emph{VAD fine-tuning}, we freeze the SSep model and only optimize the VAD parameters. Instead, in the \emph{SSep+VAD fine-tuning} method all parameters are optimized jointly.

For end-to-end trained SSGD systems we apply a modified version of the proposed leakage removal post-processing algorithm.
We found that the VAD performance could degrade when Algorithm \ref{algo:leak_rem} is applied as it is. Indeed, the VAD module is fine-tuned on the output of the SSep module, then changing the separated sources during inference introduces a mismatch with training conditions. In this case, we only employ the Algorithm \ref{algo:leak_rem} to detect leaked segments in separated estimated sources without filling them with zeros. At the end, VAD output frames associated to leaked segments are explicitly marked as non-speech frames.

\section{Experimental Setup}
\label{sec:exp_setup}

\subsection{Datasets}
\label{ssec:exp_setup/dataset}
Because the focus of our work is on the CTS scenario, we use the \emph{Fisher Corpus Part 1} and \emph{Part 2} \citep{cieri2004fisher} for both training and test purposes.
The whole Fisher consists of $11699$ telephone conversations between two participants, totaling approximately $1960$ hours of English speech sampled at 8 kHz. The amount of overlapped speech is around 14\% of the total speech duration. 
Since separated signals for the two speaker are provided, we can use this dataset to train and evaluate a separation model using common metrics as the SI-SDR improvement (SI-SDRi) \citep{leroux2019sdr}.
We split the data in $11577$, $61$ and $61$ conversations for training, validation and test sets respectively, assuring that there is no overlap between speaker identities.

In addition, we generate a simulated fully-overlapped version of Fisher for the purpose of pre-training the SSep models, as done in \cite{morrone2022conversational}. This portion is derived from the training set and amounts to 30000 mixtures for a total of 44 hours.  

To compare with state-of-the-art EEND methods, we also evaluate the proposed algorithms on the portion of the 2000 NIST SRE ~\citep{callhome} denoted as \emph{CALLHOME}. It consists of real telephone conversations in multiple languages.
We use the same 2-speaker subset of CALLHOME and the adaptation/test split proposed in \cite{fujita2019}. In this case, the amount of overlapped speech is around 13\% of the total speech duration.

Since separated sources are not available for CALLHOME, we create simulated conversations which are used to adapt SSep models to CALLHOME data. We exploit the provided annotations to extract single-speaker segments from recordings taken by the original adaptation set, discarding segments shorter than $0.1$ s. Then, the extracted segments are combined to mimic real-world conversational scenarios similar to
SparseLibriMix \citep{cosentino2020librimix}. Each conversation is created by alternately picking utterances from the two speakers, until a total minimum length of $30$ s is reached. To increase variability, we also mix speakers belonging to different recordings.
With this procedure, we generate an additional training and validation sets of $3000$ ($27.3$ h) and $500$ ($4.5$ h) examples, respectively. In this case, speakers overlap approximately $16$\% of the time.

\subsection{Architecture, Training and Inference Details}

We experiment with SSep architectures both in online and offline settings. In particular, we consider $3$ different SSep models: Conv-TasNet, DPTNet and DPRNN.
For Conv-TasNet we employ the best hyperparameter configuration proposed by \cite{luo2019conv}.
Instead, for DPRNN and DPTNet we use the hyperparameters proposed in \cite{luo2020dual} and \cite{dptnet}, respectively, with some changes. The kernel size for encoder/decoder is increased to $16$ to reduce memory consumption. The chunk and hop sizes are set to $100$ and $50$ (i.e., $50$\% overlap). In addition, the global layer normalization is replaced with standard layer normalization for online/causal models. We use the implementations available through the Asteroid toolkit \citep{pariente20}.
We pre-train SSep models on the simulated fully overlapped Fisher maximizing the SI-SDR function. We observed that the SSep pre-training allowed faster convergence compared to training from scratch on real Fisher data. 
We use Adam optimizer \citep{kingma2015adam}, batch size $4$ and learning rate $10^{-3}$. We clip gradients with $l_2$ norm greater than $5$. 
Then, we continue the training of each SSep model on $60$ s long random segments from real Fisher recording using a batch size of $1$ and reducing the learning rate to $10^{-4}$.
For Conv-TasNet and DPRNN, learning rate is halved whether SI-SDR does not improve on the validation set for $10$ consecutive epochs. For DPTNet, we employ the learning rate scheduler used in \cite{dptnet} with $k_1=1$, $k_2=4\cdot10^{-4}$, $d_{model}=64$, and $warmup\_n=5n_{epoch}$, where $n_{epoch}$ is the number of training steps performed in a single epoch. 
If no improvement is observed for $20$ epochs on validation, training is stopped.
For the CALLHOME dataset, the separators are adapted on the simulated CALLHOME using the same hyperparameters of the fine-tuned models, except for the length of training segments which is set to $30$\,s.

We employ the TCN-based causal VAD proposed by \cite{cornell2022overlapped}. The latency of the VAD is set to $0.1$ s, as the hop size of the log-Mel filterbanks input features.
This model is trained on the original Fisher Part 1, using each speaker source separately, to classify speech vs. non-speech for each input frame. We use the following weighted binary cross-entropy loss:
\begin{equation}
    \mathcal{L}_{vad}=-\frac{1}{N}\sum_{n=1}^{N} \lambda_s \cdot s_n \cdot \log(\hat{s}_n) + (1 - s_n) \cdot \log(1 - \hat{s}_n),
    \label{eq:loss_vad}
\end{equation} 
where $N$, $n$, $s_n$ and $\hat{s}_n$ are the total number of frames, the frame index, the target speech/non-speech label and the estimated speech probability, respectively. $\lambda_s$ is a parameter that is adjusted according to training data distribution. We set it to $0.9$ for all experiments.

\begin{table*}[t]
\centering
\adjustbox{max width=\textwidth}{%
\centering
\begin{tabular}{@{}l|c|c|c|ccccc@{}}
\toprule
\textbf{Method} & \textbf{VAD} & \textbf{Leakage Removal} & \textbf{Latency (s)} & \textbf{SI-SDRi} & \textbf{MS} & \textbf{FA} & \textbf{SC} & \textbf{DER} \\
\midrule
\textit{Oracle sources} & \multirow{4}{*}{Energy} & \multirow{4}{*}{\xmark} & & $\infty$ & 7.3 & 1.8 & 0.1 & 9.2 \\
Conv-TasNet & & & 0.01 & 8.9 & 9.5 & 26.9 & 1.6 & 38.0 \\
DPTNet & & & 0.1 & 21.6 & 7.5 & 2.6 & 0.8 & 10.6 \\
DPRNN & & & 0.1 & \textbf{22.7} & 7.5 & \textbf{1.7} & 0.8 & 10.0 \\
 \arrayrulecolor{black!50}\midrule
\textit{Oracle sources} & \multirow{4}{*}{TCN} & \multirow{4}{*}{\xmark} & & $\infty$ & 3.5 & 1.8 & 0.1 & 5.3 \\
Conv-TasNet & & & 0.01 & 8.9 & 7.4 & 30.9 & 5.3 & 43.6 \\
DPTNet & & & 0.1 & 21.6 & 4.3 & 3.2 & \textbf{0.6} & 8.0 \\
DPRNN & & & 0.1 & \textbf{22.7} & \textbf{3.3} & 3.5 & 0.7 & 7.5 \\
\arrayrulecolor{black!50}\midrule
Conv-TasNet & \multirow{3}{*}{TCN} & \multirow{3}{*}{\cmark} & 0.01 & 5.9 & 7.7 & 4.3 & 13.3 & 25.3 \\
DPTNet & & & 0.1 & 21.2 & 4.1 & 2.2 & 1.1 & 7.3 \\
DPRNN & & & 0.1 & 22.3 & 3.7 & 2.6 & 0.8 & \textbf{7.1} \\
 \arrayrulecolor{black}\bottomrule
\end{tabular}
}
\caption{Disjoint SSGD: speech separation and diarization results on the Fisher test set in the \textbf{online} scenario. Separation is assessed using the SI-SDR (dB) improvements over the input mixtures. Diarization is assessed using diarization error rate (DER), missed speech (MS), false alarm (FA) and speaker confusion errors (SC). Algorithmic latency is reported in seconds.
The best results among the proposed techniques are shown in \textbf{bold}.}
\label{tab:res/diar_online_fisher}
\end{table*}

The VAD is trained on $2$ s long segments and the batch size is set to $256$. We employ the same optimizer, learning rate scheduler, gradient clipping and early stopping policy used for Conv-TasNet and DPRNN.
At inference the VAD is applied on estimated separated sources independently. For each frame, the VAD predictions above a threshold $t_v$ are labeled as speech. Then, the thresholded predictions are smoothed using a median filter and segments shorter than a threshold $t_s$ are removed to mitigate false alarm errors.
The median filter length, $t_v$ and $t_s$ parameters are tuned on the Fisher validation set for each SSGD architecture. Likewise, these parameters are tuned on the CALLHOME adaptation set for CALLHOME models.
The threshold $t_{\ell r}$ and the segment length $L$ of the leakage removal algorithm are manually tuned and set to $3$ and $0.01$ s, respectively, so as not to affect the latency of the VAD.

For SSGD methods, algorithmic latencies of all modules correspond to their respective online processing unit lengths. The online processing unit is the minimum amount of buffered data needed to produce new outputs. It represents the ideal, lower bound latency that can be obtained if the computational processing times of all modules are zero. In this paper, we adopt this convention to compare latency across different diarization systems \citep{xue2021online, han2021bw, xue2021edaonline, horiguchi2022online}. For speech enhancement and separation algorithms the online processing unit often corresponds to the STFT~\citep{wang2022improving} or learned filterbank synthesis windows~\citep{luo2018tasnet, luo2019conv}. On the other hand, for separators based on the dual-path architecture (e.g., DPRNN and DPTNet) the online processing unit length is decided by the intra-chunk segment size \citep{luo2020dual, li2022skim}.
Since latency is tied to the online processing unit length, the total algorithmic latency is equal to the maximum latency among all latencies of the modules of the SSGD pipeline. However, being the modules connected in series, real-world latency on actual hardware would instead be the sum between the total algorithmic latency and the processing times sum of all modules. Actual latency times that can been obtained under specific hardware conditions are reported in Section \ref{ssec:res/rtf}.
For systems based on DPRNN and DPTNet the latency is set to $0.1$ s, which is the latency of the separators. In contrast, Conv-TasNet has lower latency than the VAD and the leakage removal algorithm, as such the resulting latency is dictated by the VAD at $0.01$ s. Full details about latency computation can be found in Appendix \ref{app:latency}.

The end-to-end models (i.e., VAD and SSep+VAD fine-tuning) are trained with the same hyperparameters above, except for the initial learning rate which is set to $10^{-5}$. The SSep module is initialized with the model trained on real Fisher and simulated CALLHOME when fine-tuned with Fisher and CALLHOME, respectively. Instead, we initialize the VAD with the model pre-trained on Fisher for both datasets. Finally, all models are fine-tuned using the real datasets to minimize the loss $\mathcal{L}_{vad}$ (cfr. Equation \ref{eq:loss_vad}).

\begin{table*}[t]
\centering
\adjustbox{max width=\textwidth}{%
\centering
\begin{tabular}{@{}l|c|c|c|cccc@{}}
\toprule
\textbf{Method} & \textbf{VAD} & \textbf{Leakage Removal} & \textbf{Latency (s)} & \textbf{MS} & \textbf{FA} & \textbf{SC} & \textbf{DER} \\
\midrule
SA-EEND w/STB~\citep{xue2021online} & \multirow{3}{*}{n.a.} & \multirow{3}{*}{n.a.} & 1 & & & & 12.5 \\
BW-EDA-EEND*~\citep{han2021bw} & & & 10 & & & &  11.8 \\
SA-EEND-EDA w/STB*~\citep{xue2021edaonline} & & & 10 & & & & 10.0 \\
EEND-GLA w/BW-STB*~\citep{horiguchi2022online} & & & 1 & & & & \underline{9.0} \\
\midrule
Conv-TasNet & \multirow{3}{*}{Energy} & \multirow{3}{*}{\xmark} & 0.01 & 7.4 & 35.3 & 3.0 & 45.7 \\
DPTNet & & & 0.1 & \textbf{5.5} & 6.3 & 0.8 & 12.6 \\
DPRNN & & & 0.1 & 5.7 & 5.2 & 1.8 & 12.7 \\
 \arrayrulecolor{black!50}\midrule
Conv-TasNet & \multirow{3}{*}{TCN} & \multirow{3}{*}{\xmark} & 0.01 & 12.2 & 36.2 & 5.1 & 53.4 \\
DPTNet & & & 0.1 & 6.6 & 4.5 & \textbf{0.6} & 11.7 \\
DPRNN & & & 0.1 & 5.9 & 3.8 & 1.7 & 11.6 \\
\arrayrulecolor{black!50}\midrule
Conv-TasNet & \multirow{3}{*}{TCN} & \multirow{3}{*}{\cmark} & 0.01 & 13.2 & 6.3 & 13.0	& 32.5 \\
DPTNet & & & 0.1 & 6.3 & 2.6 & 1.2 & \textbf{10.0} \\
DPRNN & & & 0.1 & 6.8 &	\textbf{2.2} & 2.2 & 11.2 \\
 \arrayrulecolor{black}\bottomrule
\end{tabular}
}
\caption{Disjoint SSGD: diarization results on the CALLHOME test set in the \textbf{online} scenario. Diarization is assessed using diarization error rate (DER), missed speech (MS), false alarm (FA) and speaker confusion errors (SC). Algorithmic latency is reported in seconds.
The best results among proposed techniques are shown in \textbf{bold}, and among EEND methods are \underline{underlined}.\\
\emph{*The number of speakers in input recordings is estimated by the model.}}
\label{tab:res/diar_online_callhome}
\end{table*}

\section{Experiments and Results}
\label{sec:res}

Diarization performance is evaluated in terms of diarization error rate (DER) on Fisher and CALLHOME test sets. Following \cite{fujita2019end}, we consider overlapped speech and use a start and end-point collar tolerance of $0.25$ s (i.e., \emph{fair evaluation} setup).
Moreover, we also report the DERs without collar tolerance (i.e., \emph{full evaluation} setup) obtained by end-to-end training strategies. The evaluation is carried out using the standard NIST \textit{md-eval}\footnote{\url{https://github.com/usnistgov/SCTK}} (version 22) scoring tool.
Since oracle sources are available for the Fisher test set, we also measure the separation capability using the SI-SDRi metric \citep{leroux2019sdr}.

\subsection{Online Diarization}\label{subsec:res/online_eval}

\subsubsection{Disjoint SSGD}
The results for online disjoint SSGD (i.e., SSep and VAD are trained separately) diarization models on Fisher are reported in Table~\ref{tab:res/diar_online_fisher}. Results for CALLHOME are shown in Table \ref{tab:res/diar_online_callhome}. In this case, we do not fine-tune neither SSep nor VAD on CALLHOME data. Oracle sources refer to SSGD with oracle SSep, thus with error coming only from the VAD module. Note that oracle evaluation is missing for CALLHOME, as separated sources are not available.
We also show the results of EEND on the CALLHOME test set, as reported in their respective original works. For all online EEND systems, with the exception of SA-EEND with STB \citep{xue2021online}, the results are obtained estimating the number of speakers in input recordings as these systems are only tested in this setup.

The comparison between the TCN VAD and the energy-based one highlights that, as expected, the former performs overall better. The difference however is not major, highlighting the fact that, if the separator performs well a simple VAD may be sufficient in some instances. 

The separation capability of the Conv-TasNet model was very poor, generating large false alarm errors. 
Indeed, it is limited by its short receptive fields, i.e.,  ${\sim} 1.5$ s, which prevents it to learn long term dependencies.
Such a problem is solved by the DPRNN and DPTNet which can track the speakers for much longer due to LSTMs (coupled with self-attention in DPTNet) in the inter-block. This leads to much better diarization results. These two models reached similar performance on Fisher, whereas DPTNet performed better on CALLHOME when used in conjunction with leakage removal.

Crucially, the proposed leakage removal post-processing consistently improved diarization performance for all SSep models. We observed the major benefits when leakage removal is used with the TCN-based VAD. Indeed, since it is trained on real Fisher data and not on the output of the separators (disjoint training), it is prone to classify channel leakage as active speech.
For Conv-TasNet, the DER was reduced by $42.0$\% and $39.1$\% on Fisher and CALLHOME, respectively. However, the diarization accuracy remained relatively low due to poor separation capability. For DPTNet and DPRNN, the leakage removal almost halved the false alarm errors. We observed a larger improvement with DPTNet as the DER is reduced by $8.8$\% and $14.5$\%, as opposed to DPRNN for which the DER improved by $5.3$\% and $3.4$\% on Fisher and CALLHOME, respectively.

As a comparison, the current best performing online system on the CALLHOME dataset (i.e., EEND-GLA with block-wise speaker tracing buffer \citep{horiguchi2022online}) obtains 9.0\% DER, which is slightly better than ours but is obtained with higher latency of $1$ s.
Our approach works with a latency of $0.1$ s, making it appealing for applications where strict real-time requirements are very important (e.g., real-time captioning).

\begin{table*}[t]
\centering
\adjustbox{max width=\textwidth}{%
\centering
\begin{tabular}{@{}l|c|ccccccccc@{}}
\toprule
\multirow{2}{*}{\textbf{Method}} & \multirow{2}{*}{\textbf{Online}} & \multirow{2}{*}{\textbf{SI-SDRi}} & \multicolumn{4}{c}{\textbf{Fair Eval.}} & \multicolumn{4}{c}{\textbf{Full Eval.}} \\
\cmidrule(r{4pt}){4-7} \cmidrule{8-11}
 & & & \textbf{MS} & \textbf{FA} & \textbf{SC} & \textbf{DER} & \textbf{MS} & \textbf{FA} & \textbf{SC} & \textbf{DER} \\
\midrule
\textit{Oracle sources} & n.a. & $\infty$ & 3.5 & 1.8 & 0.1 & 5.3 & 7.0 & 4.1 & 0.3 & 11.5 \\
\midrule
Disjoint SSGD w/LR & \multirow{5}{*}{\cmark} & 22.3 & 3.7 & 2.6 & 0.8 & 7.1 & 7.3 & 5.1 & 1.1 & 13.5 \\
VAD fine-tuning & & \textbf{22.7} & 3.9 & 1.6 & \textbf{0.7} & 6.2 & 7.7 & 3.6 & 0.8 & 12.1 \\
VAD fine-tuning w/LR & & \textbf{22.7} & 5.3 & \textbf{0.9} & 0.9 & 7.1 & 10.8 & \textbf{2.3} & 1.0 & 14.0 \\
SSep+VAD fine-tuning & & 14.9 & \textbf{1.9} & 1.6 & \textbf{0.7} & \textbf{4.2} & 4.3 & 3.5 & \textbf{0.7} & \textbf{8.5} \\
SSep+VAD fine-tuning w/LR & & 14.9 & 2.1 & 1.4 & \textbf{0.7} & \textbf{4.2} & \textbf{3.9} & 4.1 & \textbf{0.7} & 8.7 \\
\arrayrulecolor{black!50}\midrule
Disjoint SSGD w/LR & \multirow{5}{*}{\xmark} & 22.8 & 3.9 & 2.0 & 0.2 & 6.1 & 7.7 & 4.1 & 0.5 & 12.2 \\
VAD fine-tuning & & \textbf{23.2} & 4.0 & 1.3 & \textbf{0.1} & 5.4 & 8.0 & 3.3 & 0.3 & 11.7 \\
VAD fine-tuning w/LR & & \textbf{23.2} & 5.7 & \textbf{0.8} & \textbf{0.1} & 6.7 & 11.8 & \textbf{2.2} & 0.4 & 14.4 \\
SSep+VAD fine-tuning & & 17.7 & \textbf{3.1} & 0.9 & 0.2 & \textbf{4.1} & \textbf{5.5} & 2.7 & \textbf{0.2} & \textbf{8.4} \\
SSep+VAD fine-tuning w/LR & & 17.7 & 3.2 & \textbf{0.8} & 0.2 & 4.2 & 5.8 & 2.5 & \textbf{0.2} & 8.6 \\
\arrayrulecolor{black}\bottomrule
\end{tabular}
}
\caption{End-to-end SSGD: separation and diarization results on the Fisher test set. The best results among proposed techniques are shown in \textbf{bold}. "w/LR" is appended to method name when leakage removal is applied.}
\label{tab:res/e2e_fisher}
\end{table*}

\begin{table*}[t]
\centering
\adjustbox{max width=\textwidth}{%
\centering
\begin{tabular}{@{}l|c|ccccccccc@{}}
\toprule
\multirow{2}{*}{\textbf{Method}} & \multirow{2}{*}{\textbf{Online}} & \multicolumn{4}{c}{\textbf{Fair Eval.}} & \multicolumn{4}{c}{\textbf{Full Eval.}} \\
\cmidrule(r{4pt}){3-6} \cmidrule{7-10}
 & & \textbf{MS} & \textbf{FA} & \textbf{SC} & \textbf{DER} & \textbf{MS} & \textbf{FA} & \textbf{SC} & \textbf{DER} \\
\midrule
Disjoint SSGD w/LR & \multirow{6}{*}{\cmark} & 6.8 & 2.2 & 2.2 & 11.2 & 9.8 & 10.6 & 3.1 & 23.6 \\
Disjoint SSGD w/LR w/Simu-CH adapt. & & 6.7 & 2.2 & 0.6 & 9.5 & \textbf{6.5} & 11.2 & 1.5 & 22.1 \\
VAD fine-tuning & & 6.8 & 2.3 & 0.6 & 9.8 & 10.9 & 6.9 & 1.3 & 19.1 \\
VAD fine-tuning w/LR & & 7.6 & \textbf{1.5} & 0.8 & 9.8 & 12.5 & \textbf{5.3} & 1.5 & 19.3 \\
SSep+VAD fine-tuning & & \textbf{6.3} & 2.1 & \textbf{0.4} & \textbf{8.8} & 8.8 & 7.0 & \textbf{0.7} & \textbf{16.5} \\
SSep+VAD fine-tuning w/LR & & 6.5 & 1.9 & \textbf{0.4} & \textbf{8.8} & 9.3 & 6.5 & 0.8 & 16.6 \\
\arrayrulecolor{black!50}\midrule
Disjoint SSGD w/LR & \multirow{6}{*}{\xmark} & 6.2 & 2.6 & 1.5 & 10.2 & 8.5 & 12.1 & 2.3 & 22.9 \\
Disjoint SSGD w/LR w/Simu-CH adapt. & & \textbf{5.6} & 2.9 & 0.8 & 9.3 & \textbf{7.7} & 12.9 & 1.5 & 22.1 \\
VAD fine-tuning & & 7.2 & 2.1 & 1.0 & 10.2 & 11.3 & 6.2 & 1.6 & 19.1 \\
VAD fine-tuning w/LR & & 7.6 & \textbf{1.3} & 1.2 & 10.0 & 12.3 & \textbf{5.0} & 1.8 & 19.1 \\
SSep+VAD fine-tuning & & 6.5 & 2.0 & \textbf{0.7} & \textbf{9.2} & 9.2 & 6.2 & \textbf{1.0} & \textbf{16.4} \\
SSep+VAD fine-tuning w/LR & & 6.7 & 1.8 & \textbf{0.7} & \textbf{9.2} & 9.8 & 5.7 & \textbf{1.0} & 16.6 \\
\arrayrulecolor{black}\bottomrule
\end{tabular}
}
\caption{End-to-end SSGD: diarization results on the CALLHOME test set. The best results among proposed techniques are shown in \textbf{bold}. "w/LR" is appended to method name when leakage removal is applied. "w/Simu-CH adapt." is appended when the separator is adapted on the simulated CALLHOME data.}
\label{tab:res/e2e_callhome}
\end{table*}

\subsubsection{End-to-End SSGD}
\label{sssec:res/e2e}

We focus on the DPRNN-based architecture as it performs best on the Fisher dataset. Additionally, we observed that the adaptation of DPRNN on simulated CALLHOME is more stable compared to other separators (i.e., Conv-TasNet and DPTNet) leading to better final performance on the real CALLHOME test set.

We experiment with different adaptation and end-to-end training strategies. Contrary to what we have reported for disjoint SSGD models, we also report here the \emph{full evaluation} (no collar evaluation) as it can provide a better estimation of segmentation accuracy.

Table \ref{tab:res/e2e_fisher} reports the results on the Fisher test set. The two fine-tuning strategies provided meaningful improvements over the online disjoint SSGD with leakage removal. In particular, the VAD and SSep+VAD fine-tuning strategies reduced the \emph{fair/full} DER by $12.7$/$10.3$\% and $40.8$/$37.0$\%, respectively. Both approaches helped a lot to reduce FA errors. In these cases, the use of the leakage removal algorithm was not useful. It even degraded the performances when used with the VAD fine-tuning. Note that the SSep+VAD fine-tuning strategies also outperformed the evaluation with oracle sources, meaning that the model was able to estimate ``custom'' separated sources which resulted in better diarization outputs. The significant improvement in diarization performance was obtained at the cost of lowering separation performance which however still remains acceptable.

The results on the CALLHOME test set are shown in Table \ref{tab:res/e2e_callhome}. The adaptation of the separator on the simulated CALLHOME was very effective to reduce SC errors and then the overall DER. As such, we apply all the end-to-end strategies starting from this SSGD adapted on the simulated CALLHOME. The VAD fine-tuning only improved with the full evaluation. Instead, the SSep+VAD fine-tuning reduced the \emph{fair/full} DER by $7.3$/$24.9$\% over the adapted disjoint SSGD.

To the best of our knowledge, the SSep+VAD approach \emph{fair} evaluation outperformed all online state-of-the-art EEND methods on the 2-speaker CALLHOME test set with significantly lower latency, i.e., $0.1$ vs $1$ or $10$ s.
Additionally, the SSGD is trained using a dataset of ${\sim} 1900$ hours of speech, which is about 5x smaller  than the ones used to train the state-of-the-art EEND models (i.e., ${\sim} 10000$ hours).

\begin{table*}[t]
\centering
\adjustbox{max width=\textwidth}{%
\centering
\begin{tabular}{@{}l|c|c|ccccc@{}}
\toprule
\textbf{Method} & \textbf{VAD} & \textbf{Leakage Removal} & \textbf{SI-SDRi} & \textbf{MS} & \textbf{FA} & \textbf{SC} & \textbf{DER} \\
\midrule
VBx~\citep{landini2022bayesian} & TCN & \multirow{5}{*}{n.a.} & \multirow{5}{*}{n.a.} & 10.0 & \underline{0.3} & 0.5 & 10.8 \\
VBx~\citep{landini2022bayesian} & Kaldi & & & 8.9 & 0.4 & 0.9 & 10.2 \\
~~ + Overlap assignment~\citep{bullock2020overlap} & Kaldi & & & \underline{4.4} & 2.1 & 0.9 & \underline{7.4} \\
Spectral clustering~\citep{park2019auto} & Kaldi & & & 8.9 & 0.4 & \underline{0.2} & 9.5 \\
~~ + Overlap assignment~\citep{raj2021multi} & Kaldi & & & 5.2 & 2.0 & \underline{0.2} & \underline{7.4} \\
\midrule
\textit{Oracle sources} & \multirow{4}{*}{Energy} & \multirow{4}{*}{\xmark} & $\infty$ & 7.3 & 1.8 & 0.1 & 9.2 \\
Conv-TasNet & & & 20.1 & 7.6 & 4.1 & 0.9 & 12.7 \\
DPTNet & & & 22.2 & 7.5 & 2.1 & 0.4 & 10.0 \\
DPRNN & & & \textbf{23.2} & 7.5 & 1.6 & 0.2 & 9.3 \\
 \arrayrulecolor{black!50}\midrule
\textit{Oracle sources} & \multirow{4}{*}{TCN} & \multirow{4}{*}{\xmark} & $\infty$ & 3.5 & 1.8 & 0.1 & 5.3 \\
Conv-TasNet & & & 20.1 & 4.1 & 5.2 & 0.7 & 10.0 \\
DPTNet & & & 22.2 & 4.8 & 2.4 & \textbf{0.1} & 7.3 \\
DPRNN & & & \textbf{23.2} & \textbf{3.4} & 3.1 & \textbf{0.1} & 6.5 \\
\arrayrulecolor{black!50}\midrule
Conv-TasNet & \multirow{3}{*}{TCN} & \multirow{3}{*}{\cmark} & 19.7 & 4.9 & \textbf{1.3} & 1.6 & 7.9 \\
DPTNet & & & 21.9 & 5.5 & 1.4 & \textbf{0.1} & 7.0 \\
DPRNN & & & 22.8 & 3.9 & 2.0 & 0.2 & \textbf{6.1} \\
 \arrayrulecolor{black}\bottomrule
\end{tabular}
}
\caption{Disjoint SSGD: speech separation and diarization results on the Fisher test set in the \textbf{offline} scenario.
The best results among proposed techniques are shown in \textbf{bold}, and among baselines are \underline{underlined}.}
\label{tab:res/diar_offline_fisher}
\end{table*}

\begin{table*}[t]
\centering
\adjustbox{max width=\textwidth}{%
\centering
\begin{tabular}{@{}l|c|c|cccc@{}}
\toprule
\textbf{Method} & \textbf{VAD} & \textbf{Leakage Removal} & \textbf{MS} & \textbf{FA} & \textbf{SC} & \textbf{DER} \\
\midrule
VBx~\citep{landini2022bayesian} & TCN & \multirow{9}{*}{n.a.} & 7.3 & 1.9 & 3.1 & 12.3 \\
VBx~\citep{landini2022bayesian} & Kaldi & & 8.3 & \underline{0.9} & 2.6 & 11.7 \\
~~ + Overlap assignment~\citep{bullock2020overlap} & Kaldi & & 5.3 & 2.5 & 2.4 & 10.3 \\
Spectral clustering~\citep{park2019auto} & Kaldi &  & 8.3 & \underline{0.9} & 5.3 & 14.5 \\
~~ + Overlap assignment~\citep{raj2021multi} & Kaldi & & 5.7 & 2.7 & 5.8 & 14.1 \\
SA-EEND~\citep{fujita2019end} & n.a. & & & & & 9.5 \\
SA-EEND-EDA*~\citep{horiguchi2020end} & n.a. & & & & & 8.1 \\
EEND-VC*~\citep{kinoshita21_interspeech} & n.a. & & \underline{4.0} & 2.4 & \underline{0.5} & 7.0 \\
EEND-GLA*~\citep{horiguchi2021towards} & n.a. & & & & & 6.9 \\
DIVE~\citep{zeghidour2021dive} & n.a. & & & & & \underline{6.7} \\
\midrule
Conv-TasNet & \multirow{3}{*}{Energy} & \multirow{3}{*}{\xmark} & 5.5 & 4.7	& 0.7 & 10.9 \\
DPTNet & & & \textbf{5.4} & 5.3 & \textbf{0.4} & 11.1 \\
DPRNN & & & 5.5 & 5.2 & 0.9 & 11.6 \\
 \arrayrulecolor{black!50}\midrule
Conv-TasNet & \multirow{3}{*}{TCN} & \multirow{3}{*}{\xmark} & 6.5 & 4.4 & 0.5 & 11.4 \\
DPTNet & & & 7.0 & 3.3 & \textbf{0.4} & 10.7 \\
DPRNN & & & 6.5 & 4.0 & 0.7 & 11.2 \\
\arrayrulecolor{black!50}\midrule
Conv-TasNet & \multirow{3}{*}{TCN} & \multirow{3}{*}{\cmark} & 6.1 & 2.6 & 0.9 & 9.6 \\
DPTNet & & & 6.3 & \textbf{2.4} & 0.8 & \textbf{9.5} \\
DPRNN & & & 6.2 & 2.6 & 1.5 & 10.2 \\
 \arrayrulecolor{black}\bottomrule
\end{tabular}
}
\caption{Disjoint SSGD: diarization results on the CALLHOME test set in the \textbf{offline} scenario. The best results among proposed techniques are shown in \textbf{bold}, and among baselines/EEND methods are \underline{underlined}.\\
\emph{*The number of speakers in input recordings is estimated by the model.}}
\label{tab:res/diar_offline_callhome}
\end{table*}

\subsection{Offline Diarization}
\label{subsec:res/offline_eval}

We compare the offline SSGD with both clustering-based and EEND methods.
As clustering-based baselines we use VBx~\citep{landini2022bayesian} and spectral clustering~\citep{park2019auto}, along with their overlap-aware counterparts \citep{bullock2020overlap, raj2021multi}. In this case, we use the publicly available Kaldi ASpIRE VAD ~\citep{peddinti2015jhu} model\footnote{\url{https://kaldi-asr.org/models/m4}}. For overlap detection, we fine-tune the Pyannote~\citep{bredin2020pyannote} segmentation model\footnote{\url{https://huggingface.co/pyannote/segmentation}} on the full CALLHOME adaptation set. We tune the hyperparameters for each aforementioned module on the associated validation sets. To perform a fair comparison, we also tested the use of the TCN-based VAD in the VBx system, which however led to lower performance. For CALLHOME, we also report diarization errors of state-of-the-art EEND systems, as done in Table \ref{tab:res/diar_online_callhome}. For all the clustering-based baselines we assume that the oracle number of speakers (i.e., 2) is known in order to perform a fair comparison with the proposed SSGD approach. Regarding EEND approaches, for SA-EEND \citep{fujita2019end} and DIVE \citep{zeghidour2021dive} the maximum number of speakers is known, while for the other systems the speaker counting is estimated by the model. In particular, although the oracle number of speakers can be provided for EEND-VC, we decide to report the performance with estimated speaker counting as it leads to the best results in the 2-speaker scenario as reported in \cite{kinoshita21_interspeech}.

The results for the offline setting are reported in Tables~\ref{tab:res/diar_offline_fisher} and \ref{tab:res/diar_offline_callhome} for Fisher and CALLHOME, respectively.

Regarding separation performance (SI-SDRi), we can see that the offline DPTNet and DPRNN slightly outperformed their online counterparts.
The offline Conv-TasNet was able to obtain good separation capability, resulting in diarization performances similar the DPTNet and DPRNN for CALLHOME. Here, the use of non-causal convolutions and global layer normalization allows the model to track correctly the speakers. However, the dual-path SSep methods outperformed Conv-TasNet on the Fisher test set on all metrics. The overlap-aware VBx resulted the best among the clustering-based baselines, which were all surpassed by our best SSGD systems on the two test sets.
Regarding separation performance (SI-SDRi), the offline DPRNN and DPTNet slightly outperformed their online counterparts which confirmed that both dual-path SSep online models are very effective when only past context is available.

As in the online setting, the TCN VAD outperformed the energy-based one and the proposed leakage removal algorithm continued to be useful.

Regarding the proposed adaptation and end-to-end training methods, we can make similar considerations with respect to the online scenario (cfr. Section \ref{sssec:res/e2e}).

For the CALLHOME data, the best performing offline model is comparable with SA-EEND \citep{fujita2019end}, although it is not competitive with the current best performing approaches \citep{kinoshita2021integrating, zeghidour2021dive, horiguchi2021towards}, making it less attractive for offline applications. 
However, it can be a cost-effective solution as the separated signals can be readily used in downstream applications such as ASR. We will show this in Section \ref{subsec:res/wer_eval}.

\begin{figure*}[ht]
\centering
    \begin{subfigure}[b]{0.45\textwidth}
         \includegraphics[width=\textwidth]{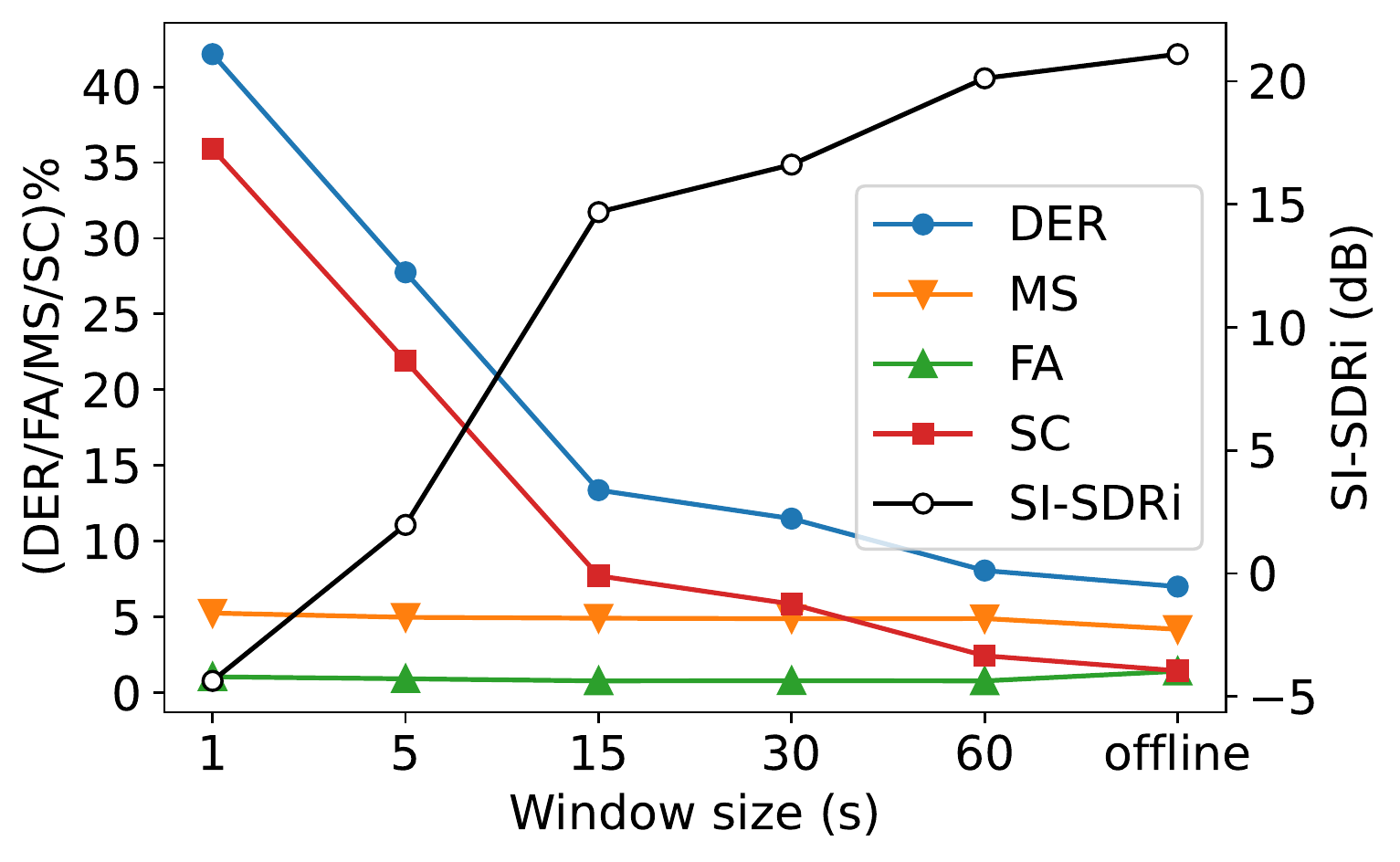}
         \caption{Fisher}
         \label{fig:css_analysis/fisher}
     \end{subfigure}
     \begin{subfigure}[b]{0.41\textwidth}
         \includegraphics[width=\textwidth]{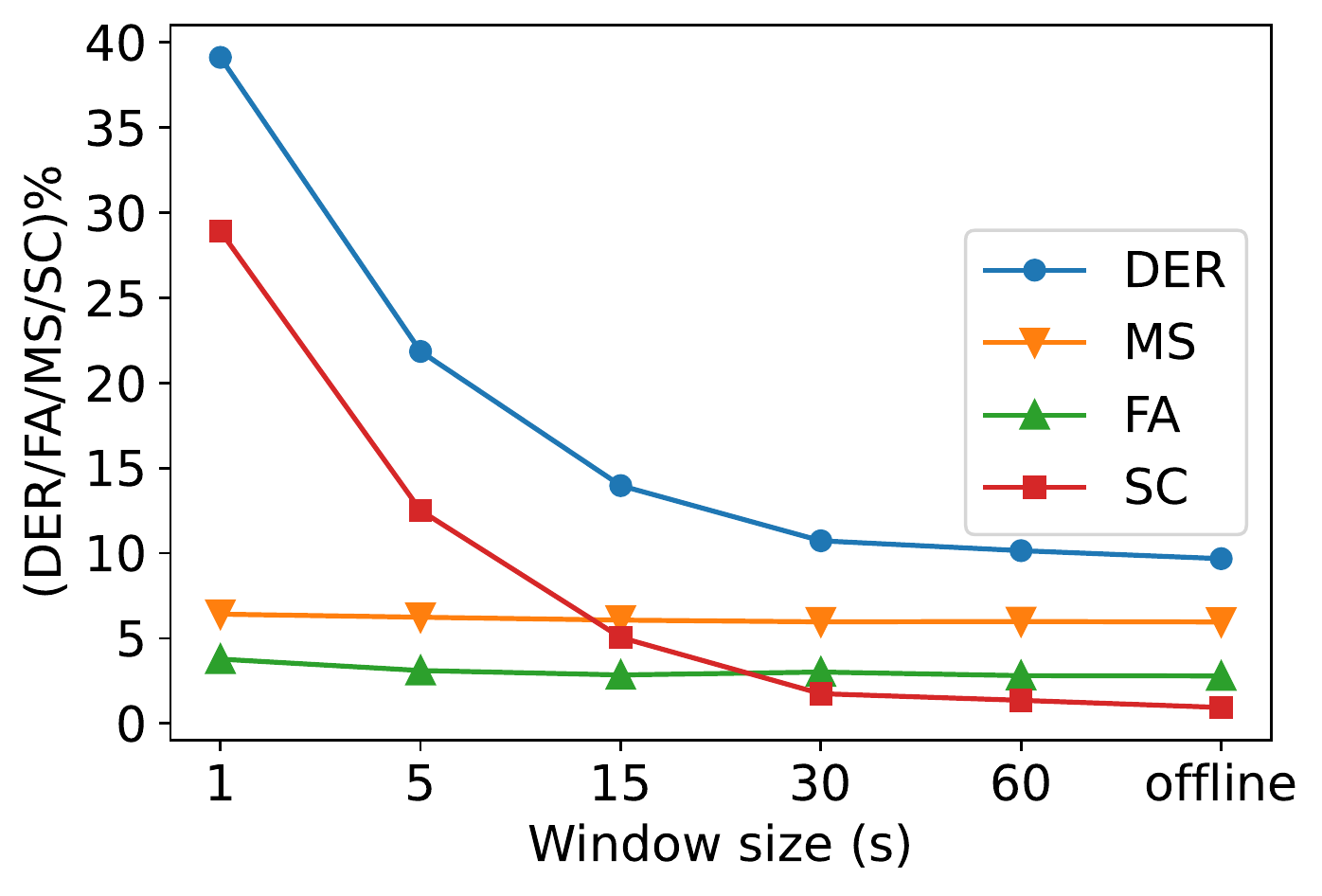}
         \caption{CALLHOME}
         \label{fig:css_analysis/callhome}
     \end{subfigure}
     \caption{Separation and diarization results on the test sets with different CSS windows. The overlap between windows is set to $50$\%. The results are obtained with the disjoint SSGD model (DPRNN+TCN+Leakage removal).}
     \label{fig:css_analysis}
\end{figure*}

\subsection{CSS Window Analysis}
\label{subsec:res/css_win_eval}

The CSS framework allows the processing of arbitrarily long inputs using chunk-wise processing. We can also exploit CSS to reduce latency of a non-causal SSep model to the CSS window length. In this way, we implement an alternative approach to perform online diarization which employs an offline SSep model in an SSGD framework.

For our purposes, we consider the offline DPRNN-based disjoint SSGD with leakage removal from Table \ref{tab:res/diar_offline_fisher} to analyze how varying CSS windows size affects diarization performance.

\begin{figure*}[ht]
\centering
    \begin{subfigure}[b]{0.45\textwidth}
         \includegraphics[width=\textwidth]{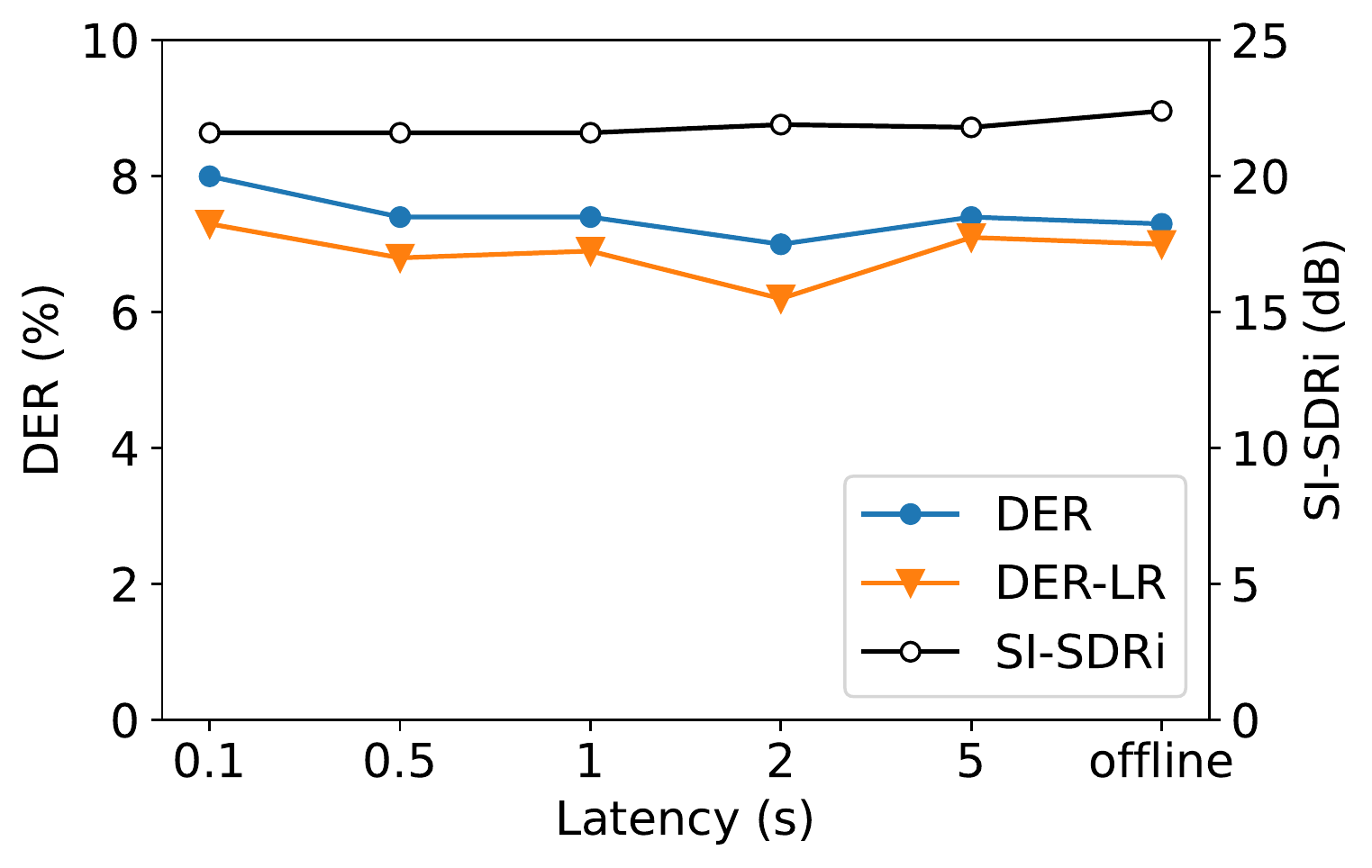}
         \caption{Fisher}
         \label{fig:latency_analysis/fisher}
     \end{subfigure}
     \begin{subfigure}[b]{0.41\textwidth}
         \includegraphics[width=\textwidth]{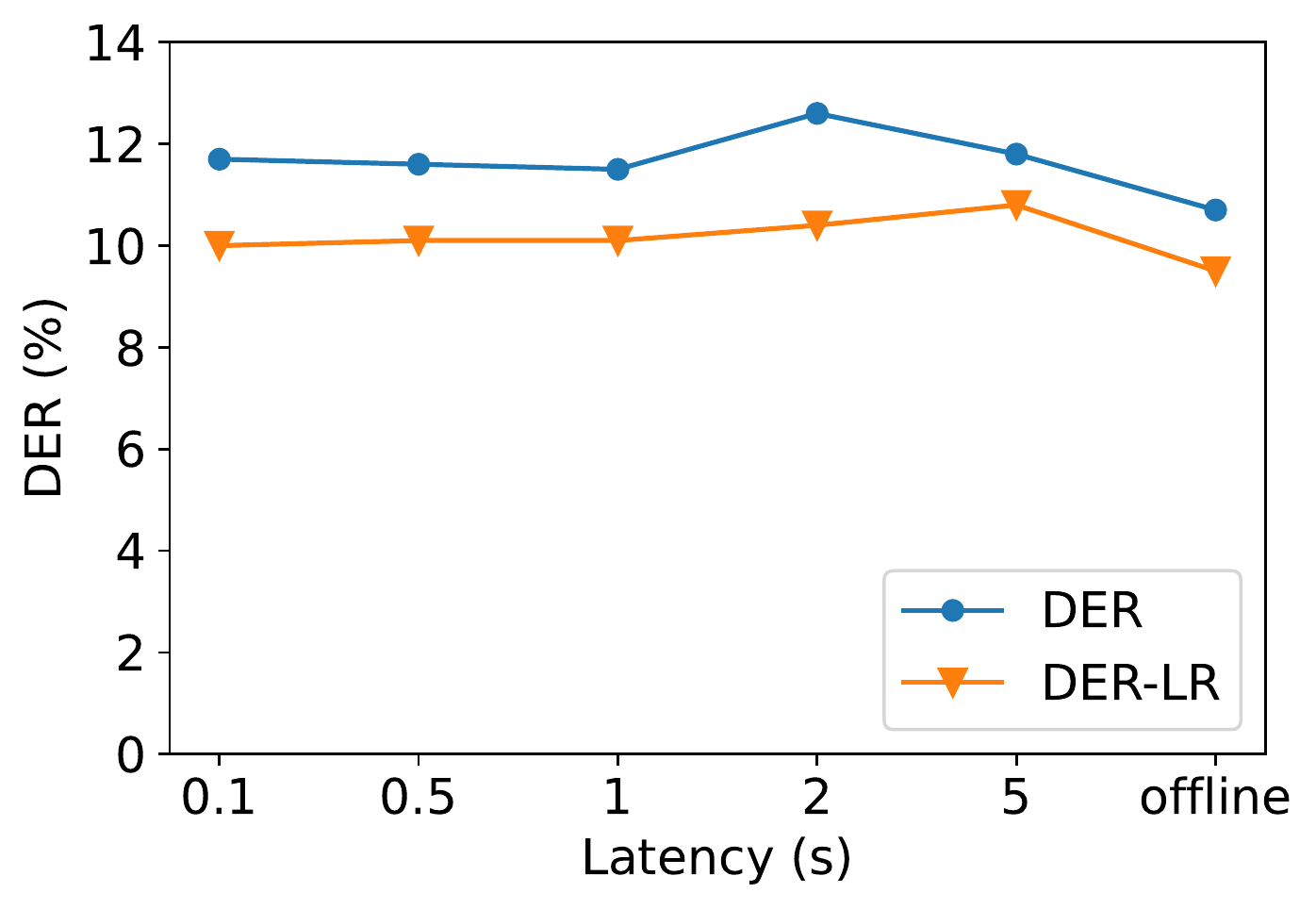}
         \caption{CALLHOME}
         \label{fig:latency_analysis/callhome}
     \end{subfigure}
     \caption{Separation and diarization results on the test sets with the online DPTNet-based SSGD by varying latency.}
     \label{fig:latency_analysis}
\end{figure*}

The results are shown in Fig.~\ref{fig:css_analysis} for both datasets
Generally, we observed that the main source of error came from speaker confusion which consistently decreased for larger CSS windows, while missed speech and false alarm error rates remained approximately constant. 
In the presence of longer input recordings the CSS benefited from using longer windows to reduce errors in computing the correct permutation across frames during the stitching stage. Indeed, the correct permutation is computed on the overlapped portion using the cross-correlation, which is more reliable when computed on larger segments.
The performances were close to the offline ones for windows larger than $60$ and $30$ s for Fisher and CALLHOME, respectively. The different slope of DER curves in Fig.~\ref{fig:css_analysis/fisher} and~\ref{fig:css_analysis/callhome} is due to the different average recording duration, which is $10$ minutes and $72$ s for Fisher and CALLHOME, respectively. 
This finding also suggests a parallelization strategy for offline CSS. Indeed, using shorter windows (e.g., $30$ or $60$ s) results in a lower memory footprint and higher inference speed-ups without affecting separation and diarization capabilities. 

Compared to the online SSGD approach with causal SSep (Sec. \ref{subsec:res/online_eval}), the CSS framework requires higher latency to obtain the same performance (i.e. $0.1$ vs $30$/$60$ s). On the other hand, it could be an appropriate option in applications in which a lower memory footprint and better ASR accuracy are important requirements rather than low latency, especially for very long recordings (e.g., $>10$ minutes).

\subsection{Latency Analysis with DPTNet}
We employ the online DPTNet-based SSGD to study how diarization performance changes by varying model latency from $0.1$ to $5$ s using the inter-chunk self-attention layers. In this way, the additional latency allows the separator to exploit more future context which may improve separation and diarization capabilities. We do not experiment with the online DPRNN-based model as it is not possible to modify latency while keeping the same hyper-parameters (e.g., intra-chunk size) thus performing a fair analysis.
The results are shown in Fig. \ref{fig:latency_analysis}. For both Fisher and CALLHOME test sets, there is not a clear trend meaning that DPTNet is not able to take much advantage of more future information at least till $5$~s lookahead. 
For this reason, it is always convenient to use the model with the lowest latency (i.e., $0.1$ s) or the full offline system when streaming processing is not required.

\begin{figure*}[ht]
\centering
    \begin{subfigure}[b]{0.48\textwidth}
         \includegraphics[width=\textwidth]{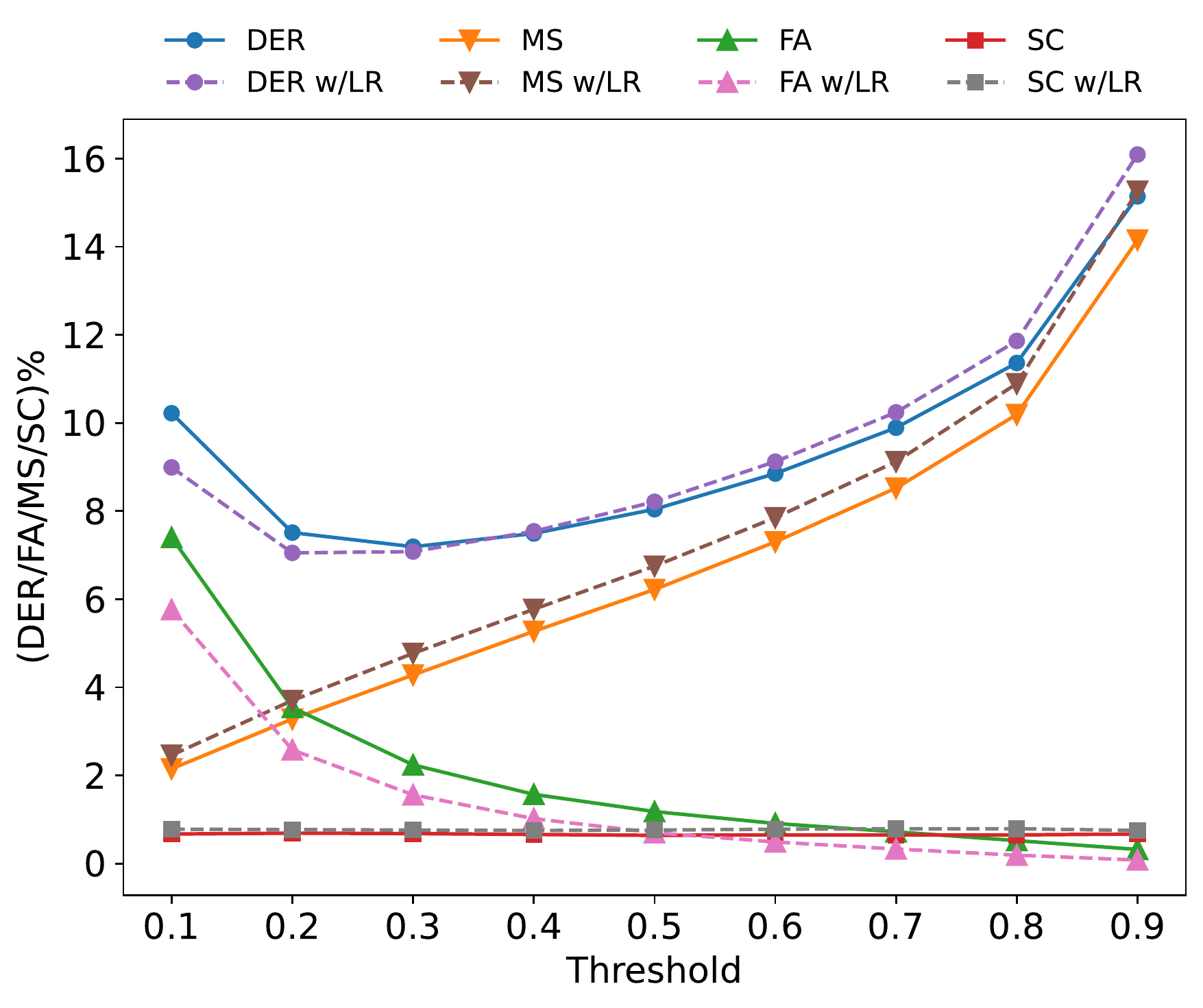}
         \caption{Fisher}
         \label{fig:ssgd_thresh_analysis/fisher}
     \end{subfigure}
     \begin{subfigure}[b]{0.48\textwidth}
         \includegraphics[width=\textwidth]{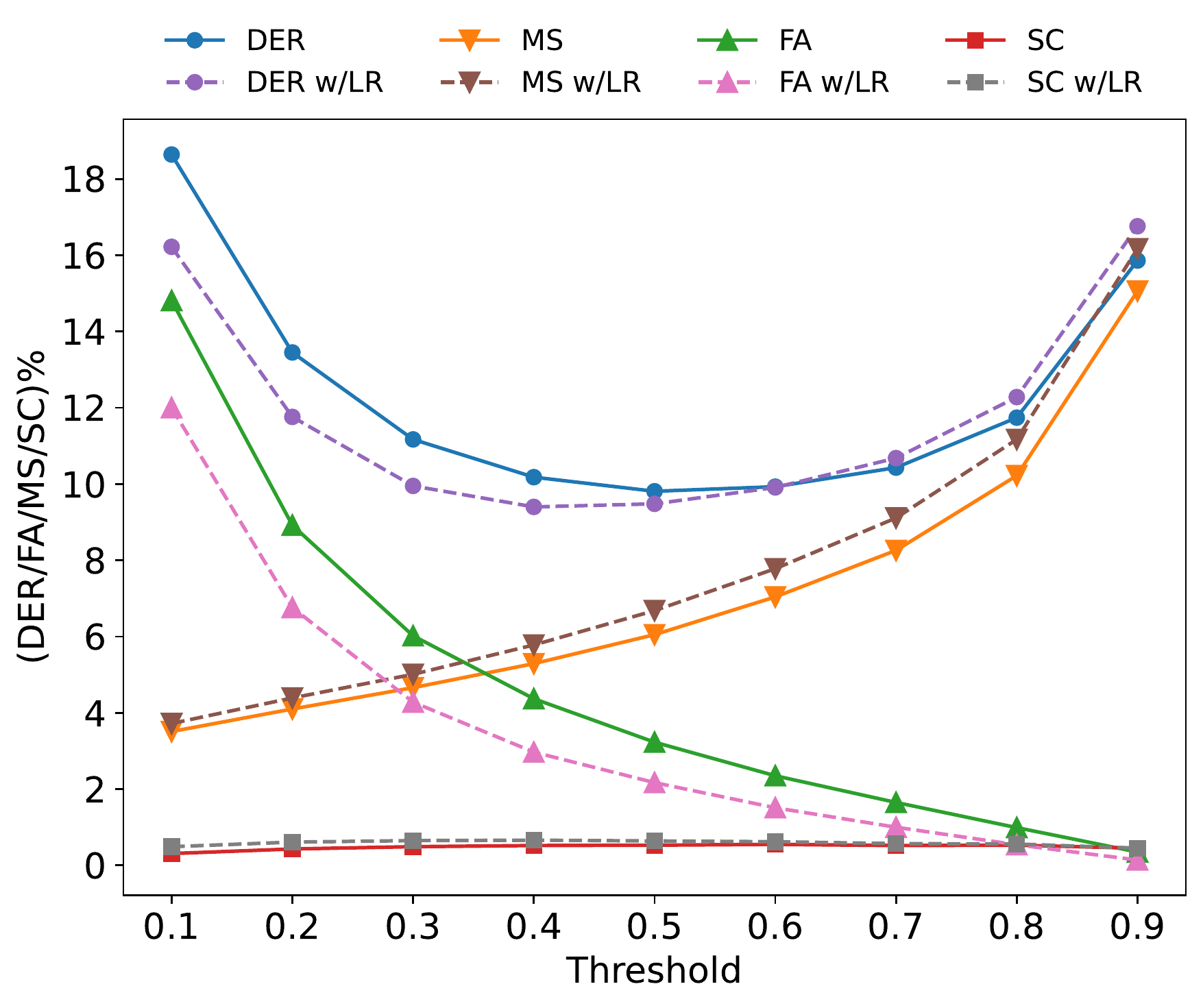}
         \caption{CALLHOME}
         \label{fig:ssgd_thresh_analysis/callhome}
     \end{subfigure}
     \caption{Diarization results of the disjoint SSGD with and without leakage removal on the test sets by varying VAD threshold.}
     \label{fig:ssgd_thresh_analysis}
\end{figure*}

\begin{figure*}[ht]
\centering
    \begin{subfigure}[b]{0.48\textwidth}
         \includegraphics[width=\textwidth]{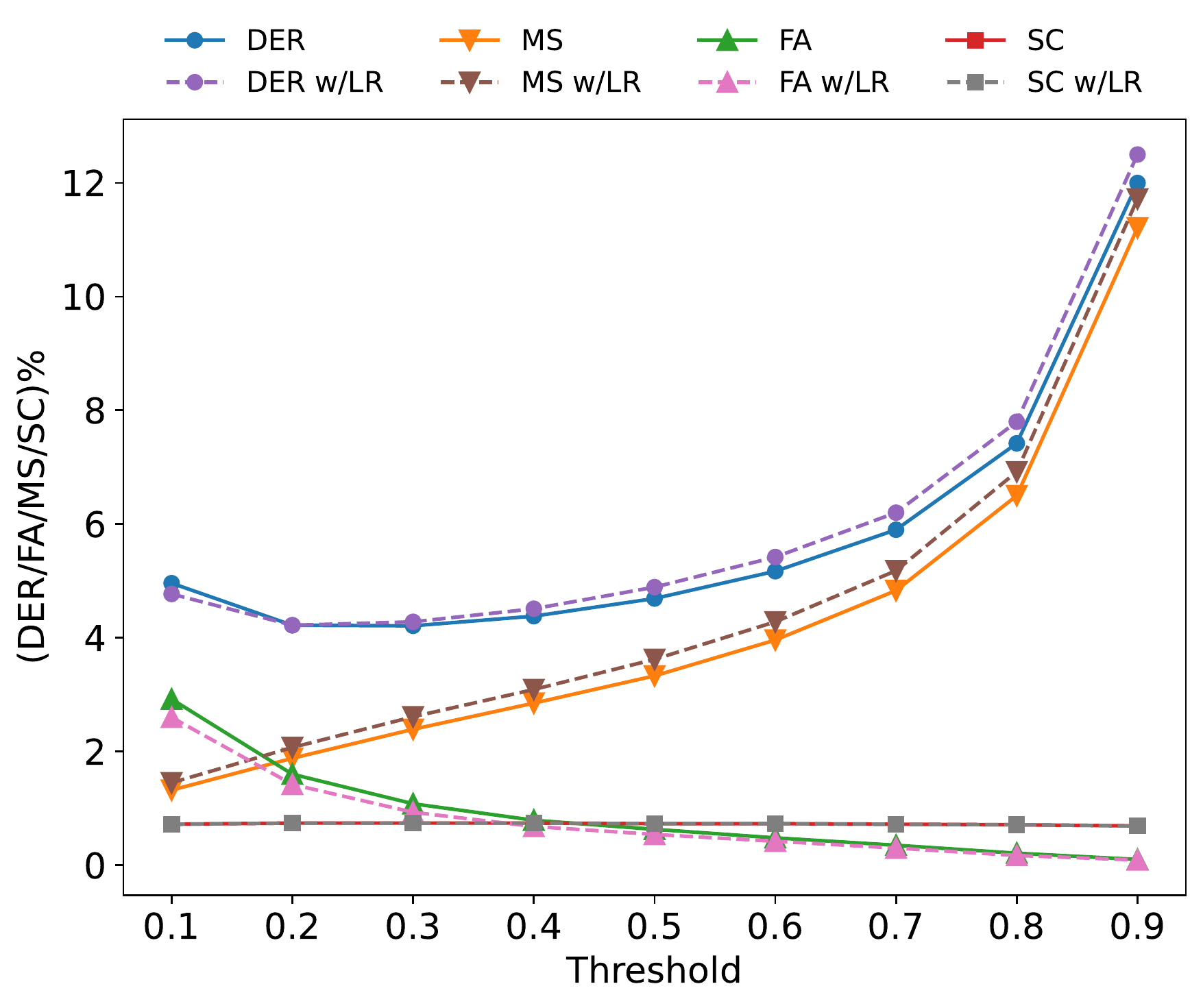}
         \caption{Fisher}
         \label{fig:e2e_ssgd_thresh_analysis/fisher}
     \end{subfigure}
     \begin{subfigure}[b]{0.48\textwidth}
         \includegraphics[width=\textwidth]{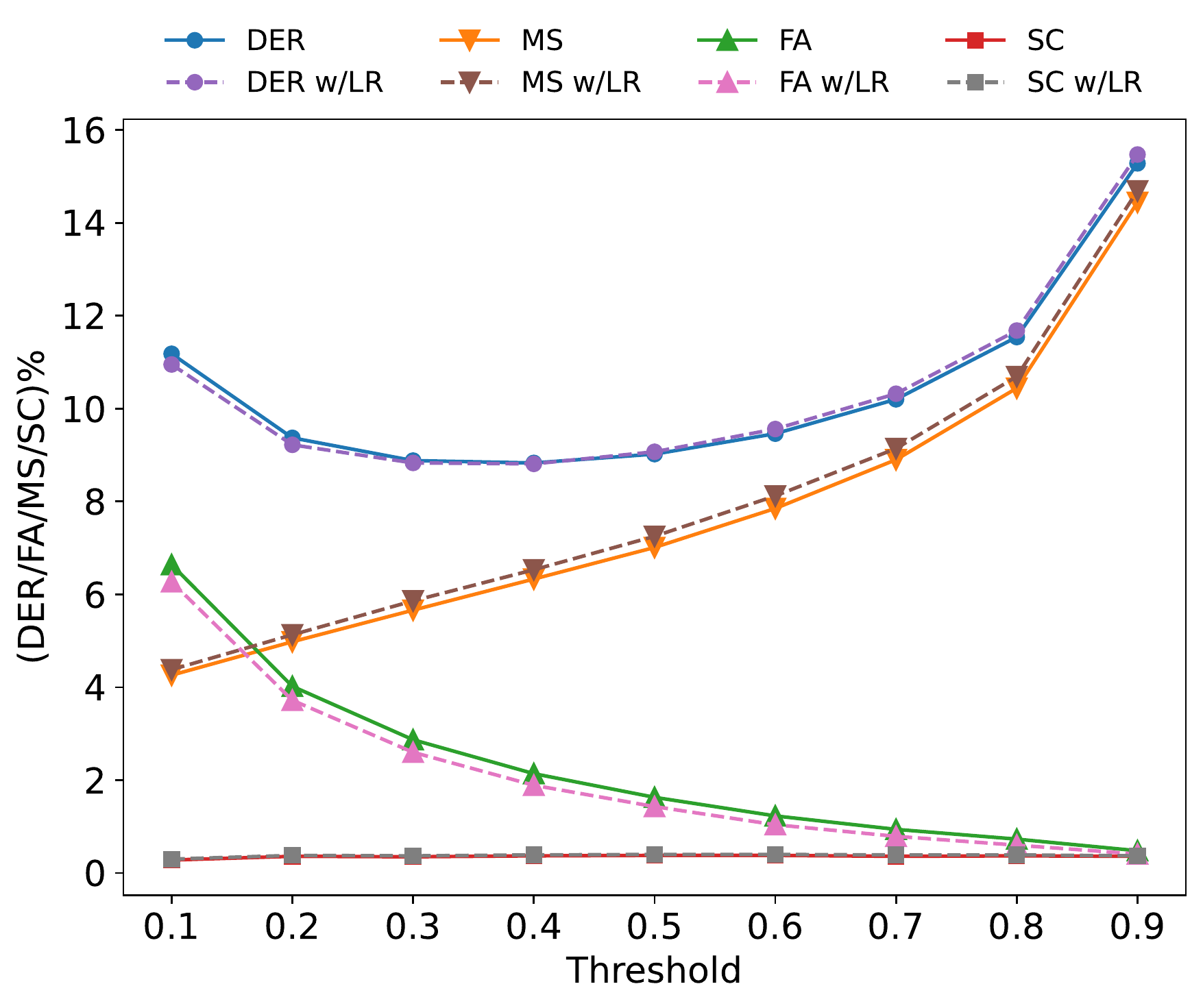}
         \caption{CALLHOME}
         \label{fig:e2e_ssgd_thresh_analysis/callhome}
     \end{subfigure}
     \caption{Diarization results of the end-to-end SSGD with and without leakage removal on the test sets by varying VAD threshold.}
     \label{fig:e2e_ssgd_thresh_analysis}
\end{figure*}

\begin{figure*}[ht]
\centering
    \begin{subfigure}[b]{0.5\textwidth}
         \includegraphics[width=\textwidth]{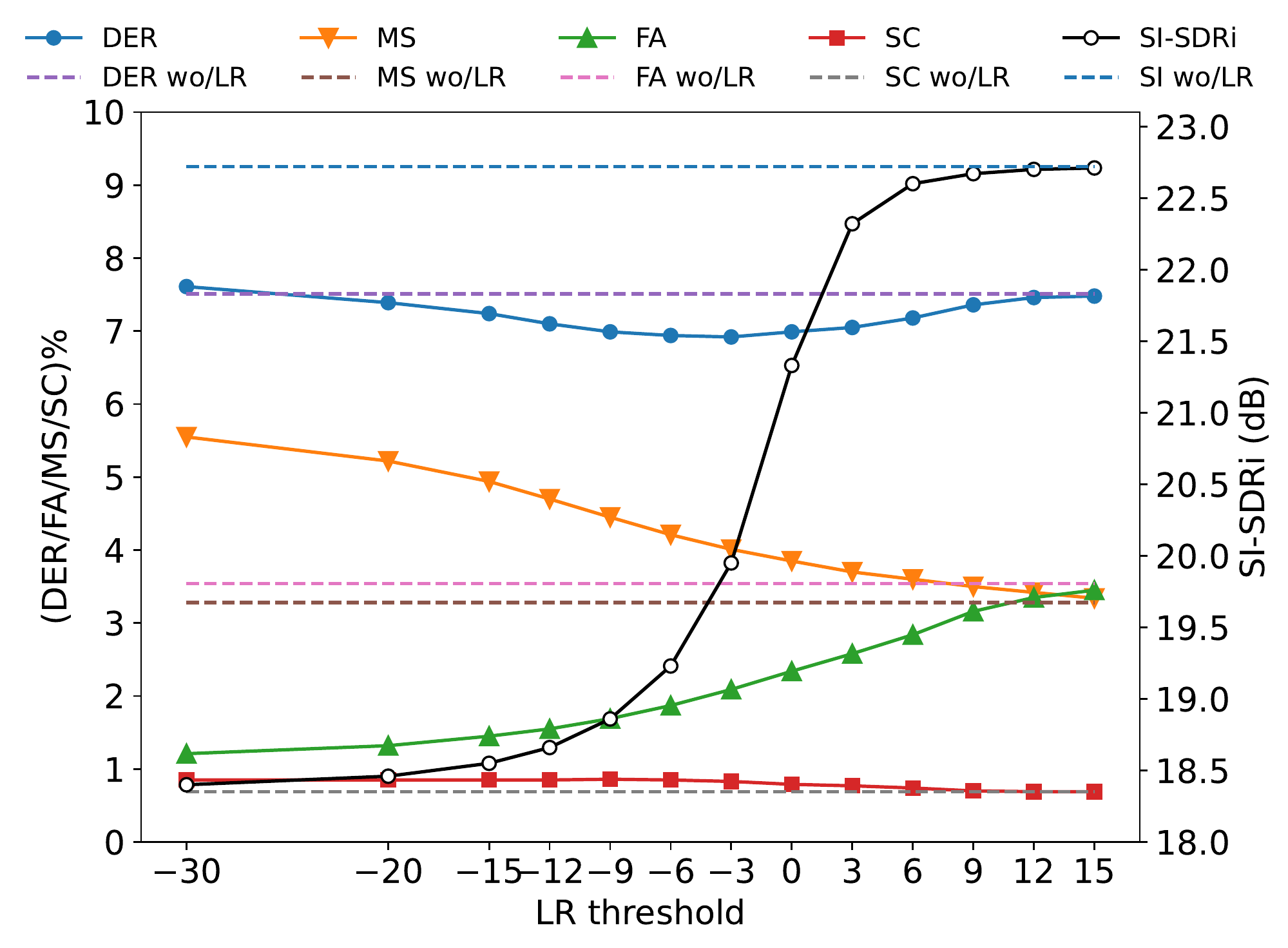}
         \caption{Fisher}
         \label{fig:ssgd_lr_thresh_analysis/fisher}
     \end{subfigure}
     \begin{subfigure}[b]{0.44\textwidth}
         \includegraphics[width=\textwidth]{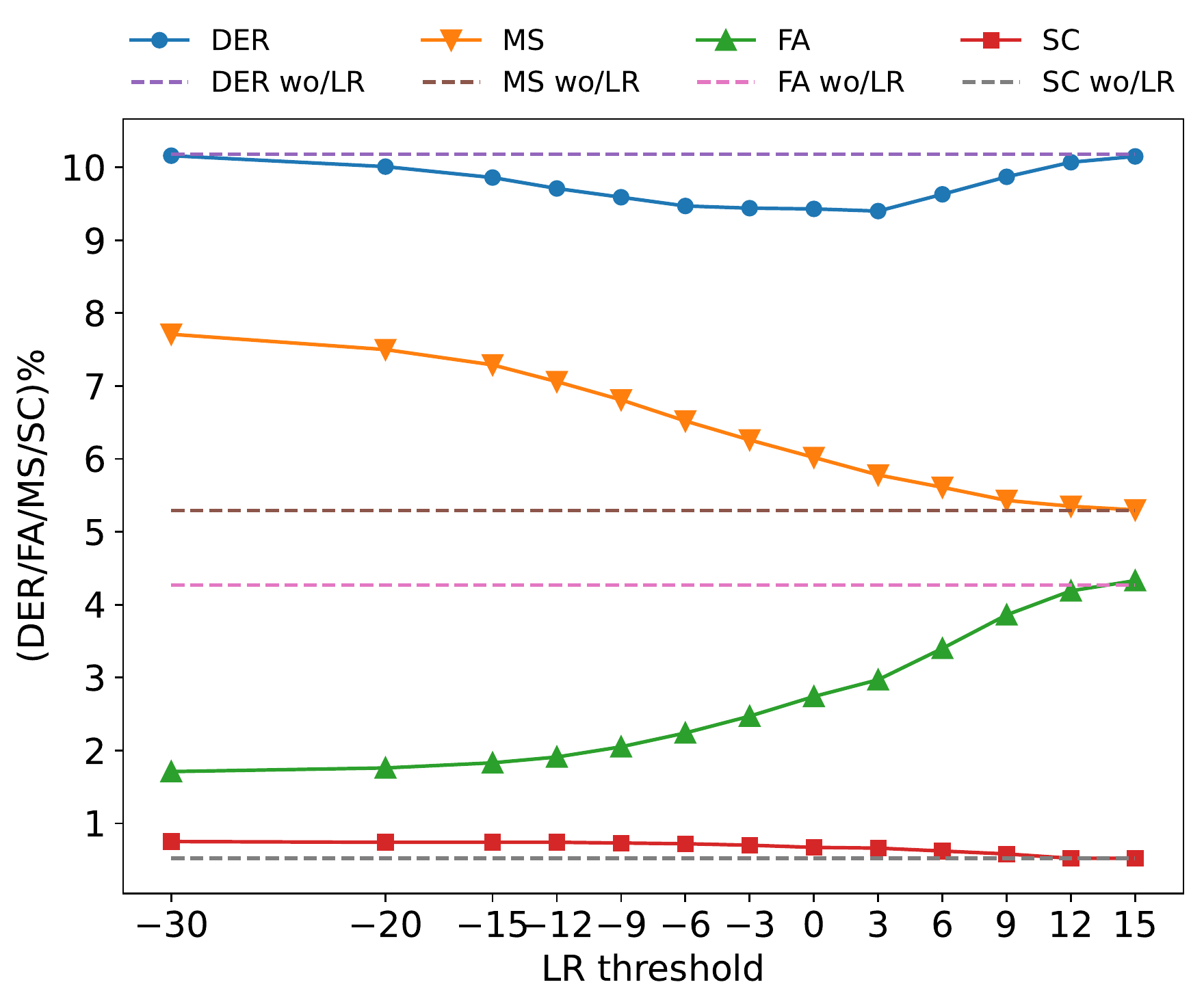}
         \caption{CALLHOME}
         \label{fig:ssgd_lr_thresh_analysis/callhome}
     \end{subfigure}
     \caption{Diarization results of the disjoint SSGD with leakage removal on the test sets by varying leakage removal (LR) threshold $t_{\ell r}$.}
     \label{fig:ssgd_lr_thresh_analysis}
\end{figure*}

\subsection{Leakage Removal Analysis}
\label{subsec:leak_rem}
We perform a study in order to analyze the effect of the leakage removal algorithm on disjoint and end-to-end SSGD. In particular, we employ the same DPRNN-based model by varying the VAD threshold $t_v$.

Fig. \ref{fig:ssgd_thresh_analysis} shows the results of the disjoint SSGD on the Fisher and CALLHOME datasets. As expected, lower thresholds generated higher FA and lower MS errors, whilst higher thresholds produced the opposite behavior (i.e., lower FA and higher MS). On the other hand, SC remained roughly constant. The use of leakage removal consistently reduced DER and FA when the VAD threshold is below a value depending on the dataset. When the threshold is above that value, FA errors are very close to zero and the higher degradation of MS resulted in higher DER. On the CALLHOME dataset, the leakage removal algorithm was way more effective compared to Fisher. Indeed, the VAD is only trained on Fisher data, and leakage removal was able to mitigate the mismatch between the data distributions of the two datasets.

The results of the end-to-end SSGD are depicted in Fig. \ref{fig:e2e_ssgd_thresh_analysis}. Contrary to the disjoint SSGD, the leakage removal did not improve overall diarization performance not even for low thresholds. This demonstrated that the end-to-end training was able to mitigate channel leakage as the VAD is updated with the separator output. In this case, the use of leakage removal would be a convenient choice only when having lower FA is desirable.

Additionally, we investigate how varying the leakage removal threshold, $t_{\ell r}$, affects the DERs. The results are reported in Fig. \ref{fig:ssgd_lr_thresh_analysis}. Both datasets showed similar trends. When $t_{\ell r}$ is very low the leakage removal algorithm greatly reduced false alarms introducing additional missed speech at the same time. On the other hand, for sufficiently high $t_{\ell r}$ values, the algorithm did not change the output and the metrics are on par with the one obtained by models without leakage removal. For all thresholds in between the overall performance was generally improved. The best DERs were obtained setting $t_{\ell r}$ to $-3$ and $3$, respectively. However, the best threshold for the DER metric highly degraded the SI-SDR on the Fisher test set. Then, we set the threshold $t_{\ell r}$ to $3$ for all experiments to achieve an optimal balance between separation and diarization capabilities.

\subsection{ASR Evaluation}
\label{subsec:res/wer_eval}

The SSGD framework outputs both separated sources and segmentation which can be readily fed in input to a back-end ASR system. It is a great advantage over other diarization methods.
We investigate the ASR performance when estimated sources from DPRNN models are fed to a downstream ASR, considering or not leakage removal. We also analyze the effect of the TCN-based VAD compared to oracle segmentation.
We use the pre-trained Kaldi ASPiRE ASR \citep{povey2011kaldi} model~\footnote{\url{https://kaldi-asr.org/models/m1}} and report the performance in terms of word error rate (WER).
Additionally, we also report the results obtained with input mixture and oracle sources, which represent the upper and lower bound for ASR evaluation. 

The results are reported in Table \ref{tab:res/wer_eval}.
The degradation of all SSGD systems, except for SSep+VAD fine-tuning, was very small compared to the evaluation with oracle sources. Additionally, the SSGD obtained large improvements over the mixtures. These findings confirm the effectiveness of the proposed SSep methods.
In general, the significant gap between the proposed model with the fully oracle system (i.e., oracle VAD + oracle sources) demonstrated that the VAD segmentation represents the main source of error. This finding is consistent with diarization results where the errors mainly came from MS and FA.

\begin{table}[t]
\centering
\adjustbox{max width=\textwidth}{%
\centering
\begin{tabular}{@{}l|c|ccc@{}}
\toprule
\multirow{2}{*}{\textbf{Method}} &
\multirow{2}{*}{\textbf{Online}} &
\multicolumn{2}{c}{\textbf{VAD}} \\
\cmidrule(r{4pt}){3-4}
 &  & \multicolumn{1}{l}{\textbf{TCN}} & \multicolumn{1}{l}{\textbf{Oracle}}\\
\midrule
\textit{Mixture} & \multirow{2}{*}{n.a.} & 38.74  & 30.69 & \\
\textit{Oracle sources} & & 25.44 &  19.50 &  \\
\midrule
DPRNN  & \multirow{5}{*}{\cmark} & 26.28 & \textbf{20.63}  & \\
+ Leakage removal & & 26.67  & 21.12 & \\
+ Leakage removal (seg-only) &   & 26.23 & \textbf{20.63} & \\
+ VAD fine-tuning & & \textbf{26.04} & \textbf{20.63} & \\
+ SSep+VAD fine-tuning & & 31.82 & 29.29 & \\
\midrule
DPRNN & \multirow{5}{*}{\xmark} & 25.77 &  \textbf{19.98} & \\
+ Leakage removal &  & 26.22  & 20.34 &  \\
+ Leakage removal (seg-only) &   & 25.79 & \textbf{19.98} & \\
+ VAD fine-tuning &  & \textbf{25.65} & \textbf{19.98} & \\
+ SSep+VAD fine-tuning & &   27.48 & 21.79 & \\
\bottomrule
\end{tabular}
}
\caption{WER evaluation on the Fisher test set. The best online/offline non-oracle results are reported in \textbf{bold}.}
\label{tab:res/wer_eval}
\end{table}

Although the leakage removal post-processing generally improves diarization, the filled zeros could produce fewer natural utterances which negatively affect the WER. On the other hand, in the proposed framework it could be only exploited for obtaining the segmentation using the non-processed estimated sources (\emph{+ Leakage removal (seg-only)}). In this latter case the DER is slightly reduced, as well for the VAD fine-tuning. Indeed, better segmentation improves performance when used with the same separated sources.
The lower separation capability of models (cfr. Table \ref{tab:res/e2e_fisher}) trained with the SSep+VAD fine-tuning resulted in higher WER as, during fine-tuning, there are no guarantees the output of the separator will be distortion-free, and even subtle distortions, while minimal in terms of SI-SDR, are known to affect ASR models significantly \citep{von2020end}. 
Adding distortion-free constraints while fine-tuning, such as multi-frame minimum variance distortion-less response \citep{tammen2021deep} is a possible future direction.

\subsection{Real Time Factor and Latency}
\label{ssec:res/rtf}

To show that the proposed methods can be run in a truly online manner, we calculated the real time factor (RTF) of online DPRNN-based SSGD with and without the leakage removal module. The online processing unit length was of $0.1$ s, which corresponds to the algorithmic latency of online DPRNN-based SSGD. The RTF is computed as the ratio between the processing time and the length of an online processing unit. Since the end-to-end training strategies only change the model parameters without affecting the system architecture, the reported results are valid for both disjoint and end-to-end trained SSGD. We carried out our experiments on a Intel\textsuperscript{\textregistered} Core\textsuperscript{\texttrademark} i9-10920X CPU @ 3.50 GHz using one thread without any GPU. The RTF was equal to $0.159$ and $0.161$ for systems with and without leakage removal, respectively. As a result, the average latency time was about $0.116$ s for all DPRNN-based systems. This demonstrates that the proposed approach is applicable for real-time inference.

\section{Conclusion and Future Work}
\label{sec:conc}

In this paper, we have explored end-to-end integration of SSGD components, i.e. the speech separator and VAD. We have focused on low-latency models which allow diarization for arbitrarily long inputs in a frame-online fashion. The proposed fine-tuning strategies showed significant improvements on both Fisher and CALLHOME datasets compared to the system without fine-tuning. In particular, our best online model, i.e., SSep+VAD fine-tuning, outperformed the current state-of-the-art methods based on EEND on the CALLHOME dataset with an order of magnitude lower latency (i.e, $0.1$ s vs. $1$ s).

Additionally, we have extended our previous work by performing a more comprehensive analysis of SSGD in real-world telephone conversation scenarios.
We have experimented with another SSep model, i.e., DPTNet, which can be easily adapted to perform online inference with desired latency. Experimental results demonstrated that additional latency did not improve diarization accuracy, thus it was always convenient to use the model with the lowest latency.
We have also generated a simulated dataset for the purpose of adaptation of SSep models to CALLHOME data. Adaptation on such data provided significant improvements, especially for DPRNN.
The use of the proposed leakage removal algorithm was investigated both for disjoint and SSep+VAD fine-tuned SSGD. Experiments clearly showed that the post-processing algorithm was very effective to reduce both FA and DER in disjoint SSGD. However, when used in conjunction with fine-tuned models it did not improve the overall performance as the models learned to handle leaked segments directly from data.
Finally, we have demonstrated that estimated sources generated by SSGD could be readily fed in input to an ASR system. The ASR performance largely improved over the input mixture and in some instances was even close to the one obtained with oracle sources.
However, the end-to-end integration led to significant ASR performance degradation. 

In future works, the SSGD framework will be extended to domains in which a higher number of speakers is typically involved (e.g., meeting scenarios).
This will need likely the development of new techniques for speech separation. One of the major issues of having more than two speakers is that the speaker tracking for arbitrarily long inputs and more speakers is significantly more difficult, potentially leading to speaker confusion errors.
Following~\cite{coria2021overlap} where a local EEND is used in conjunction with online clustering, a potential solution could be an hybrid framework in which local speech separation is followed, in the same fashion, by online clustering to achieve global speaker tracing. In such framework input signals can be split in short chunks (e.g., $5$/$10$ seconds), for which we may assume that a limited number of speakers is involved (e.g., 2 or 3). For each chunk, separated sources can be estimated jointly with speaker embeddings. Finally, online clustering can be applied on speaker embeddings to merge separated sources belonging to the same speaker across chunks. In this case, online clustering needs to be carefully designed to guarantee good inference speed without significant loss in accuracy (e.g. as done in \cite{coria2021overlap} and \cite{zhang22_odyssey}).
Moreover, the channel leakage problem also becomes more challenging as the number of output streams grows (e.g., $4$ channels for a maximum of $4$ speakers). This latter could be likely addressed by adopting separation methods that can handle a variable number of output channels~\citep{takahashi2019recursive, nachmani2020voice}.
Alternatively, our leakage removal algorithm can also be extended to deal with more output channels, by comparing each output with the other ones. Given $n$ output channels, the number of comparisons would grow as $n\,(n-1)\,/\,2$, but due to the trivial operations involved, it would still be computationally inexpensive with respect to the rest of the pipeline.
Finally, novel strategies should be investigated to mitigate the separation/ASR degradation when using end-to-end fine-tuned SSGD models. This includes distortion-less constraints and/or continual learning approaches. 

\section{Acknowledgements}
\label{sec:ack}
This work has been supported by the AGEVOLA project (SIME code 2019.0227), funded by the Trento and Rovereto Bank (CARITRO) Foundation.

\bibliographystyle{cas-model2-names}

\bibliography{refs}



\pagebreak

\onecolumn

\appendix

\section{System Latency Computation}
\label{app:latency}

In this work, we use the expression ``algorithmic latency'' $l_{algo}$ to denote the latency obtained in an ideal setting in which processing time is equal to zero. It can be seen as the lower bound for a given algorithm. In particular, algorithmic latency is tied to length of online processing unit, which is the minimum amount of buffered data needed to produce new outputs. Latencies reported for all systems in Tables \ref{tab:res/diar_online_fisher} and \ref{tab:res/diar_online_callhome} follow this convention. In real-world setups, real latency $l_{real}$ also depends on hardware conditions. In this case, since the modules are connected in series, it is correct stating that we need to sum the algorithmic latency with the processing times of all modules (i.e., $t_{SSep}$, $t_{LR}$ and $t_{VAD}$):

\begin{equation}
     l_{real} = l_{algo} + t_{SSep} + t_{LR} + t_{VAD}\\ 
\end{equation}

\begin{figure}[h]
\centering
\includegraphics[width=0.9\textwidth]{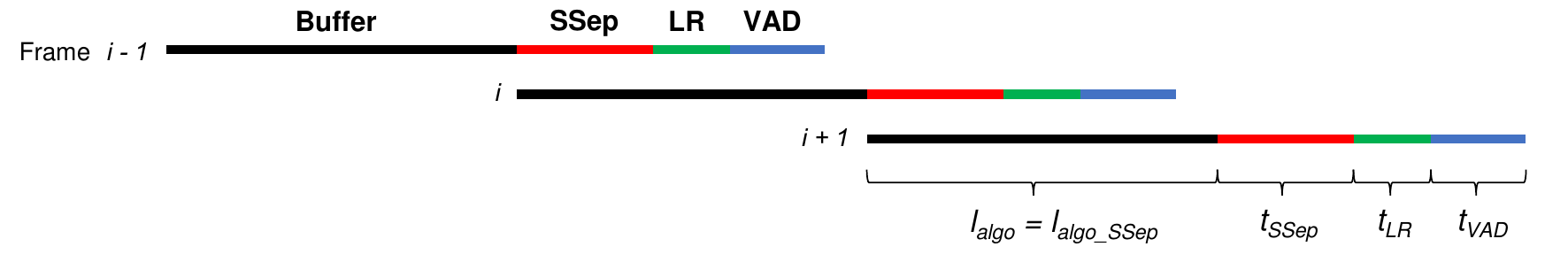}
\caption{SSGD latency diagram.}
\label{fig:latency}
\end{figure}

Below we provide a proof that the algorithmic latency of our system is equal to the largest latency among all modules. Fig. \ref{fig:latency} shows a diagram of system latency computation in consecutive frames (which correspond to online processing units).

In the online DPRNN-based SSGD, the speech separation (SSep) module has the largest algorithmic latency $l_{algo\_SSep}$. When the buffer is filled with an amount of speech equal to the SSep online processing unit length (which correspond to $l_{algo\_SSep}$), the SSep can process all the input data in the buffer. Since the algorithmic latency of the leakage removal (LR) module is lower than $l_{algo\_SSep}$, it can process all the SSep output when it is available. Then, only its processing time $t_{LR}$ contributes to the real latency $l_{real}$. The same applies to voice activity detection (VAD). Notice that during processing of frame $i-1$ the system can start to fill the buffer of next frame $i$. If the total processing time $t_{SSep} + t_{LR} + t_{VAD}$ is always lower than $l_{algo}$ (i.e., RTF $< 1$) then $l_{real}$ represents the real latency for all frames.
As a consequence, $l_{algo}=l_{algo\_{SSep}}$ and, in particular, $l_{real}=l_{algo}$ when total processing time is zero.

\end{document}